\documentclass[]{interact}

\usepackage{epstopdf}
\usepackage[caption=false]{subfig}
\usepackage[round]{natbib}

\usepackage{diagbox}
\usepackage{xcolor}
\usepackage{amsmath}
\usepackage{graphicx}
\usepackage[graphicx]{realboxes}
\usepackage{array}
\usepackage{longtable}
\usepackage{url}
\usepackage{bm}
\usepackage{multirow}
\usepackage[export]{adjustbox}
\allowdisplaybreaks
\theoremstyle{plain}
\newtheorem{theorem}{Theorem}

\theoremstyle{definition}

\theoremstyle{remark}

\begin{document}


\title{Minimum Density Power Divergence Estimation for the Generalized Exponential Distribution}

\author{
\name{Arnab Hazra \thanks{CONTACT Arnab Hazra. Email: ahazra@iitk.ac.in}}
\affil{Department of Mathematics and Statistics, Indian Institute of Technology Kanpur, Kanpur, India 208016.}
}

\maketitle

\begin{abstract}
Statistical modeling of rainfall data is an active research area in agro-meteorology. The most common models fitted to such datasets are exponential, gamma, log-normal, and Weibull distributions. As an alternative to some of these models, the generalized exponential (GE) distribution was proposed by Gupta and Kundu (2001, Exponentiated Exponential Family: An Alternative to Gamma and Weibull Distributions, Biometrical Journal). Rainfall (specifically for short periods) datasets often include outliers, and thus, a proper robust parameter estimation procedure is necessary. Here, we use the popular minimum density power divergence estimation (MDPDE) procedure developed by Basu et al. (1998, Robust and Efficient Estimation by Minimising a Density Power Divergence, Biometrika) for estimating the GE parameters. We derive the analytical expressions for the estimating equations and asymptotic distributions. We analytically compare MDPDE with maximum likelihood estimation in terms of robustness, through an influence function analysis. Besides, we study the asymptotic relative efficiency of MDPDE analytically for different parameter settings. We apply the proposed technique to some simulated datasets and two rainfall datasets from Texas, United States. The results indicate superior performance of MDPDE compared to the other existing estimation techniques in most of the scenarios.

\end{abstract}


\section{Introduction}
\label{intro}

Statistical modeling of rainfall data is a crucial research topic in agro-meteorology. Rainfall data generally exhibit positive skewness with a large upper tail. The most common probability distributions fitted to such datasets in practice are one-parameter exponential \citep{todorovic1975stochastic, burgueno2005statistical, burgueno2010statistical,hazra2018bayesian}, 
two-parameter gamma \citep{barger1949evaluation, mooley1968application, husak2007use, krishnamoorthy2014small}, two-parameter log-normal \citep{kwaku2007characterization, sharma2010use}, and two-parameter Weibull \citep{duan1995comparison, burgueno2005statistical, lana2017rainfall} distributions. While several researchers have used some goodness-of-fit tests like the chi-square test \citep{barger1949evaluation, mooley1973gamma, kwaku2007characterization}, Kolmogorov-Smirnov test \citep{sharma2010use, hazra2014modelling, al2014frequency}, variance ratio test \citep{mooley1973gamma, hazra2017note}, and Anderson-Darling test \citep{sharma2010use}, or some information criteria like the Akaike information criterion \citep{villarini2012development}, it is more common in the agro-meteorology literature to pick a reasonable probability distribution and fit it to the rainfall data.

Alternatively, in the lifetime data analysis literature, two very common probability models are three-parameter gamma and three-parameter Weibull distributions \citep{jackson1969fitting}. The three parameters controlling the location, scale, and shape allow high flexibility to capture the large upper tail behavior usually seen in the rainfall data as well. However, as pointed out by \cite{gupta1999theory}, these distributions have a limitation due to not having a closed-form analytical expression of the survival or hazard functions. Further, \cite{gupta2001exponentiated} propose a new family of two-parameter distribution, namely the exponentiated exponential distribution. The corresponding distribution function is given by
\begin{equation} \label{cdf_ge}
    F_{\textrm{GE}}(x; \lambda, \nu) = (1 - \exp[-\lambda x])^\nu; ~~~ x, \lambda, \nu > 0,
\end{equation}
where $\lambda$ and $\nu$ denote the rate and shape parameters, respectively.

While \cite{gupta1999theory} propose a more generalized version of this model by incorporating a location parameter, where the support of the distribution is parameter dependent, we stick to the form in (\ref{cdf_ge}) because rainfall data during the wet months are naturally supported over the whole positive real line and the probability of having rainfall less than a certain positive amount can possibly be negligible but not exactly zero. Further, we refer to the distribution in (\ref{cdf_ge}) as the generalized exponential (GE, henceforth) distribution. The theoretical properties of the GE distribution discussed in \cite{gupta2001exponentiated} make this model suitable for rainfall data analysis. For $\nu = 1$ in (\ref{cdf_ge}), the GE model coincides with the one-parameter exponential distribution which is among the four most popular models for rainfall data analysis \citep{hazra2017note} and thus, the GE model provides additional flexibility. The GE distribution has been used for Los Angeles rainfall data modeling by \cite{madi2007bayesian}. 

Not only in the agro-meteorology literature but also in other disciplines, the maximum likelihood estimation (MLE) is the most common parameter estimation procedure; thanks to its several attractive theoretical properties like consistency and lower asymptotic variance compared to other estimation procedures. Unfortunately, MLE gets strongly affected even in the presence of a single outlier, either a wrong data entry or an extreme observation \citep{strupczewski2005robustness, strupczewski2007robustness}. Thus, the inference about the bulk of the distribution, which is often focused in the agro-meteorology literature (except in extreme value analysis) can be erroneous. In this context, \cite{basu1998robust} propose a robust parameter estimation procedure called the minimum density power divergence estimation (MDPDE). Here we obtain the parameter estimates by minimizing a density-based divergence measure called density power divergence (DPD), over the parameter space. For a tuning parameter $\alpha\geq 0$, the DPD $d_\alpha(\cdot, \cdot)$ between two densities $f$ and $g$ is defined as
\begin{equation}\label{dpd}
    d_\alpha(g,f) = \displaystyle \left\{\begin{array}{ll}
    \displaystyle \int  \left[f^{1+\alpha}(x) - \left(1 + \alpha^{-1}\right)  f^\alpha(x) g(x) + 
\alpha^{-1} g^{1+\alpha}(x)\right] dx, & {\rm ~for} ~\alpha > 0,\\
	\displaystyle \lim_{\alpha\rightarrow 0}d_\alpha(g, f) = \int g(x) \left[\log g(x) - \log f(x) \right] dx, & {\rm ~for} ~\alpha = 0.  
\end{array}\right.
\end{equation}
Here, for $\alpha = 0$, $d_0(g, f)$ is the well-known Kullback-Leibler (KL) divergence. For parameter estimation of a certain probability model (GE, for example), the parameterized density function  $f_\theta$ plays the role of $f$ in Equation (\ref{dpd}), where $g$ denotes the true density function. Minimizing $d_0(g, f_\theta)$ returns the parameter estimates to be the same as those based on MLE. Thus, the MDPDE coincides with MLE at $\alpha=0$, and for the tuning parameter values $\alpha>0$, it provides a robust generalization of the MLE. Unlike other divergence-based approaches \citep{beran1977minimum,basu1994minimum}, MDPDE does not need any nonparametric smoothing and this allows easier implementation in practical purposes \citep{seo2017extreme}. MDPDE has been implemented in various scientific disciplines \citep{gajewski2004correspondence,yuan2008partial,seo2017extreme}, including the classical rainfall modeling literature \citep{hazra2019robust}, where MDPDE has been explored for the four most popular rainfall models described in this section.

In this paper, first, we discuss the theoretical properties of MDPDE, where we derive the asymptotic distribution of the MDPDE estimators for the parameters of the GE distribution. Further, we study the behavior of the influence function for $\lambda$ and $\nu$ at different choices of the tuning parameter $\alpha$. We then study how the fitted GE distribution varies across different $\alpha$ and describe an optimal data-driven selection of the tuning parameter $\alpha$ by minimizing the empirical Cram\'er-von Mises (CVM) distance following \cite{hazra2019robust} and \cite{fujisawa2006robust}.
Subsequently, we perform a simulation study to explore the effectiveness of MDPDE for the GE distribution at low through high levels of contamination. Here, we compare MDPDE with some alternative parameter estimation strategies like MLE, the method of moments estimation, percentile estimation, least square estimation, as well as the robust $L$-moment estimation, described in \cite{gupta2007generalized}. Further, using GE distribution and MDPDE, we analyze two rainfall datasets from Texas, United States, and compare the performance of MDPDE with several other alternatives.

The paper is organized as follows. In Section \ref{background}, we provide some background on the GE distribution, some existing parameter estimation procedures for its parameters, and the MDPDE approach. Some theoretical results along with some illustrations of the influence functions are provided in Section \ref{theory}. In Section \ref{simulation}, we perform some simulation studies to showcase the advantages of using MDPDE over other existing methods for GE model parameter estimation. We apply the statistical methodology to two rainfall datasets in Section \ref{data_application} and discuss the results. Section \ref{conclusion} concludes.

\section{Background}
\label{background}

\subsection{The generalized exponential (GE) distribution}
\label{ge_distribution}

The distribution function of the GE distribution is in (\ref{cdf_ge}). The corresponding density function is given by
\begin{equation} \label{density_ge}
    f_{\textrm{GE}}(x; \lambda, \nu) = \lambda \nu \exp[-\lambda x] (1 - \exp[-\lambda x])^{\nu - 1}; ~~~ x, \lambda, \nu > 0,
\end{equation}
and its moment-generating function is 
\begin{equation} \label{mgf_ge}
    M_{\textrm{GE}}(t; \lambda, \nu) = \Gamma(\nu + 1) \Gamma(1 - t / \lambda)/ \Gamma(\nu - t / \lambda + 1); ~~~ 0 \leq t < \lambda.
\end{equation}

\begin{figure}[h]
    \centering
\adjincludegraphics[height = 0.4\linewidth, trim = {{.0\width} {.0\width} {.0\width} {.0\width}}, clip]{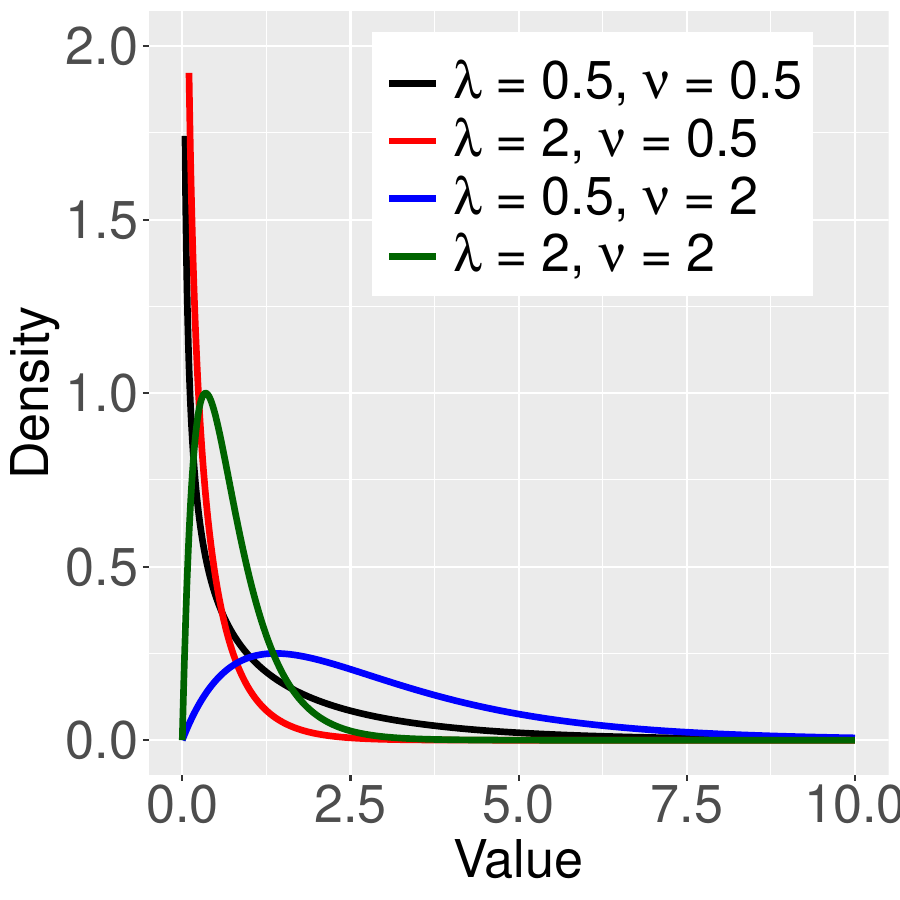}
\adjincludegraphics[height = 0.4\linewidth, trim = {{.0\width} {.0\width} {.0\width} {.0\width}}, clip]{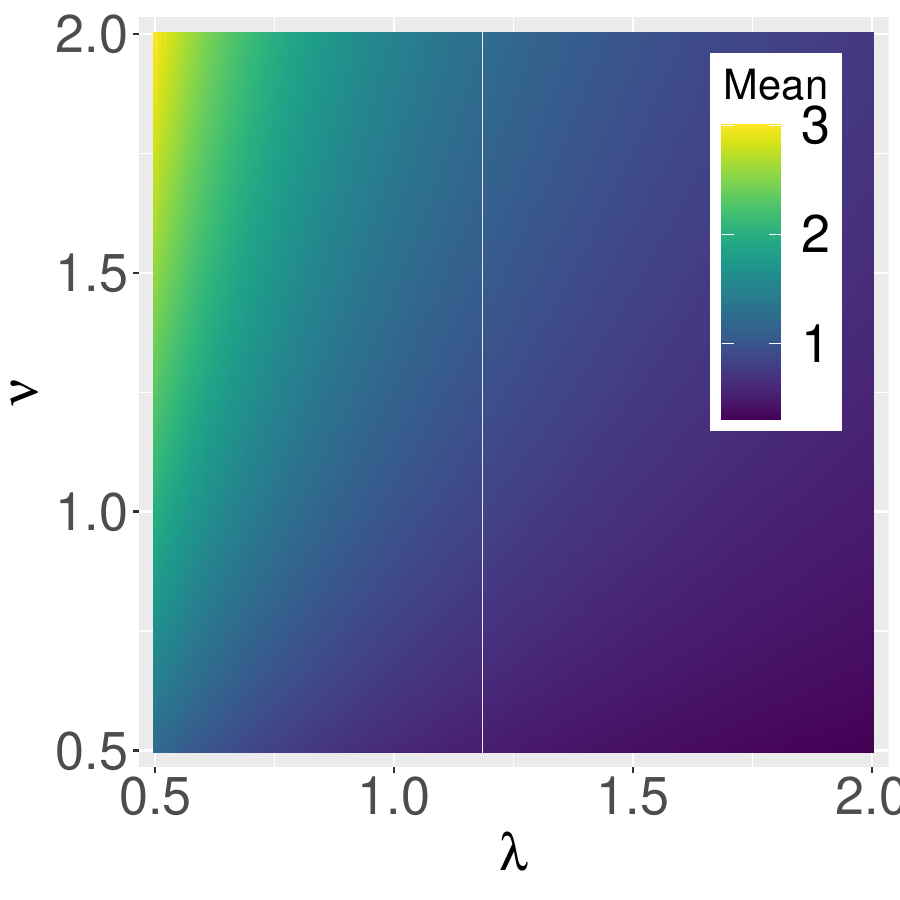} \\
\adjincludegraphics[height = 0.4\linewidth, trim = {{.0\width} {.0\width} {.0\width} {.0\width}}, clip]{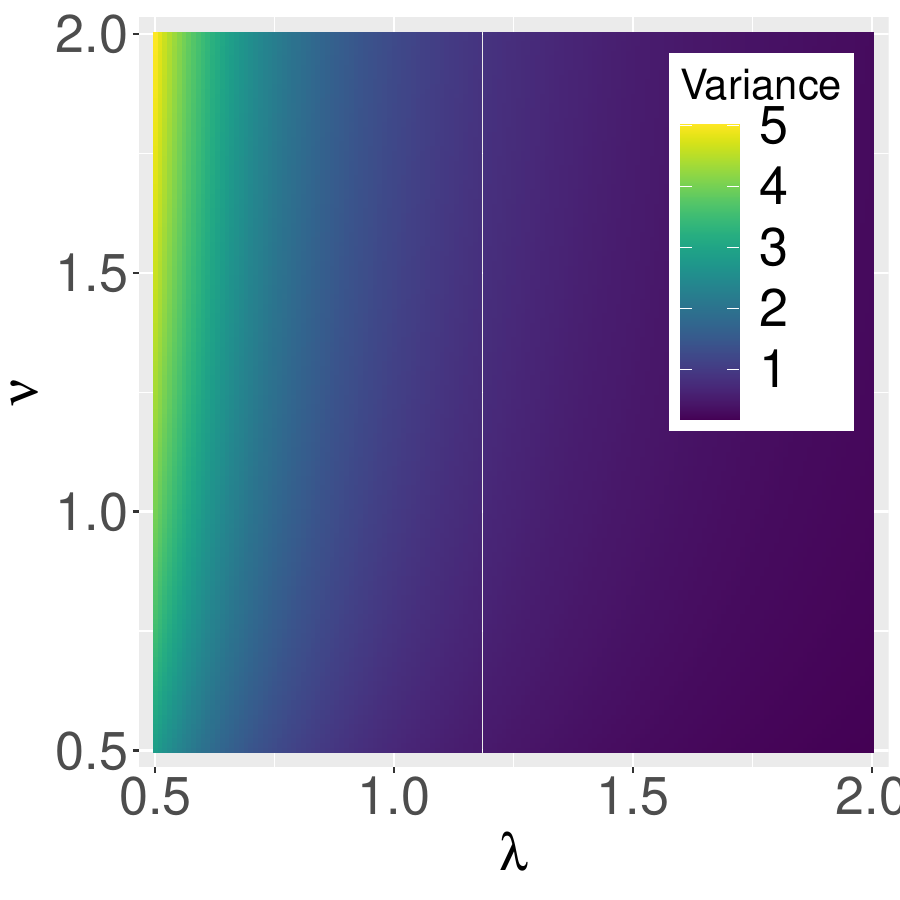}
\adjincludegraphics[height = 0.4\linewidth, trim = {{.0\width} {.0\width} {.0\width} {.0\width}}, clip]{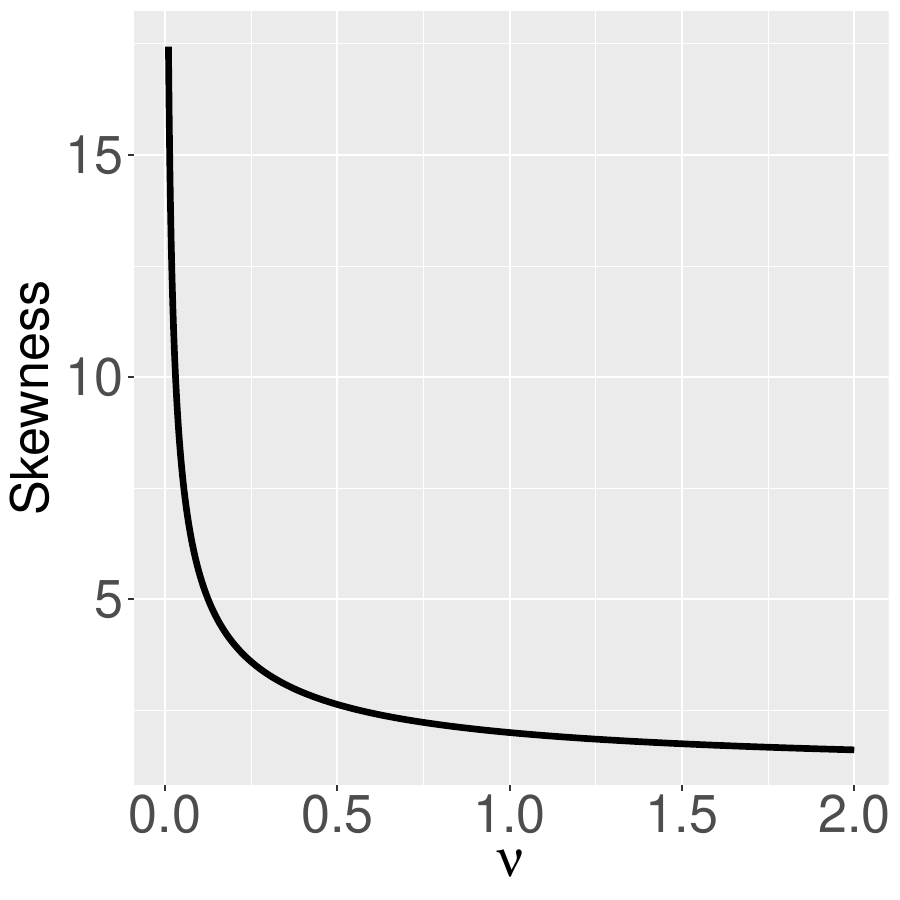}
    \caption{Top-left: GE density function for different choices of $\lambda$ and $\nu$. Top-right: GE mean function for different choices of $\lambda$ and $\nu$. Bottom-left: GE variance function for different choices of $\lambda$ and $\nu$. Bottom-right: GE skewness function (independent of $\lambda$) for different choices of $\nu$.}
    \label{fig_ge_properties}
\end{figure}

By differentiating $\log(M_{\textrm{GE}})$ iteratively and evaluating at $t=0$, the theoretical expressions of mean, variance, and skewness of a GE distributed random variable $X$ can be obtained and they are
\begin{eqnarray} \label{moments_ge}
\nonumber E(X) &=& (\psi(\nu + 1) - \psi(1)) / \lambda, \\
\nonumber Var(X) &=& (\psi'(1) - \psi'(\nu + 1)) / \lambda^2, \\
Skewness(X) &=& (\psi^{''}(\nu + 1) - \psi^{''}(1)) / (\psi^{'}(1) - \psi^{'}(\nu + 1))^{3/2},
\end{eqnarray}
where $\psi(\cdot)$, $\psi'(\cdot)$, and $\psi^{''}(\cdot)$ are polygamma functions of orders 0 (digamma function), 1 and 2, respectively. The skewness of the GE distribution does not depend on the rate parameter $\lambda$. For different values of $\lambda$ and $\nu$, the GE density, mean, variance, and skewness are illustrated in Figure \ref{fig_ge_properties}. When $\nu < 1$, the GE density function is infinite at the origin and decreases monotonically across the positive real line. If $\nu = 1$, the GE density coincides with the exponential density and finite (same as $\lambda$) at the origin and again monotonically decreases across the positive real line. When $\nu > 1$, the density function is zero at the origin, increases until its unique mode $\log(\nu) / \lambda$, and then drops towards zero. For a fixed $\nu$, the mean and variance decrease with $\lambda$, whereas for a fixed $\lambda$, the mean and variance increase with $\nu$. For $\nu$ close to zero, the GE distribution allows high positive skewness, and the skewness decreases monotonically as $\nu$ increases. By varying $\nu$, the GE model allows a more flexible skewness structure than exponential distribution, and thus it is a potentially flexible model for rainfall data analysis.

\subsection{Parameter estimation for the GE distribution}
\label{ge_parest}

There are several approaches for estimating the parameters of the GE distribution as described in \cite{gupta2007generalized}. The most common approach is the maximum likelihood (ML) estimation, where given the observed data $\mathcal{X} = \{ X_1, \ldots, X_n \}$, we assume that the observations are independent and identically distributed (IID), and we maximize the log-likelihood function $l(\lambda, \nu; \mathcal{X}) = \sum_{i=1}^n \log(f_{\textrm{GE}}(X_i; \lambda, \nu))$ over the parameter space. While no closed form expression of the estimators exists, given the ML estimator $\hat{\lambda}_{\textrm{ML}}$ for $\lambda$, the ML estimator for $\nu$ is $\hat{\nu}_{\textrm{ML}} = -n / \sum_{i=1}^n \log(1 - \exp[-\hat{\lambda}_{\textrm{ML}} X_i])$. Thus, a profile likelihood approach can be implemented in this case.

In the Method of Moments (MM) estimation, we calculate the sample mean and sample variance and equate them to the theoretical expressions for the mean and variance of the GE distribution given in (\ref{moments_ge}). From (\ref{moments_ge}), we observe that the coefficient of variation (ratio of standard deviation and mean) is independent of $\lambda$ and thus, the MM estimator of $\nu$ ($\hat{\nu}_{\textrm{MM}}$, say) can be obtained by numerically solving the equation 
$$ \mbox{SD}(\mathcal{X}) / \Bar{X} = (\psi'(1) - \psi'(\nu + 1))^{1/2} / (\psi(\nu + 1) - \psi(1)),$$
where $\Bar{X}$ and $\mbox{SD}(\mathcal{X})$ are the sample mean and sample standard deviation obtained from $\mathcal{X}$. Further, we have $\hat{\lambda}_{\textrm{MM}} = (\psi(\hat{\nu}_{\textrm{MM}} + 1) - \psi(1)) / \Bar{X}$.

In the Percentile (PT) estimation, we minimize the sum of the squared distance between the ordered statistics and the theoretical quantile function of the GE distribution given by
\begin{equation} \label{quantile_ge}
    Q_{\textrm{GE}}(p; \lambda, \nu) = -\lambda^{-1} \log(1 - p^{1 / \nu}); ~~~ p \in (0, 1), \lambda, \nu > 0,
\end{equation}
where the parameters $\lambda$ and $\nu$ have the same interpretation as in (\ref{cdf_ge}). Suppose, the ordered observations in $\mathcal{X}$ be denoted by $X_{(1)} < \ldots < X_{(n)}$. Then, the empirical distribution function at $X_{(i)}$ is (approximately) $i / (n + 1)$ for each $i=1, \ldots, n$. Thus, $\hat{\lambda}_{\textrm{PT}}$ and $\hat{\nu}_{\textrm{PT}}$, the percentile estimators of $\lambda$ and $\nu$, are obtained by minimizing $\sum_{i=1}^n (X_{(i)} - Q_{\textrm{GE}}(i / (n + 1); \lambda, \nu))^2$.

In the Least Square (LS) estimation, we obtain $\hat{\lambda}_{\textrm{LS}}$ and $\hat{\nu}_{\textrm{LS}}$, the estimators of $\lambda$ and $\nu$, by minimizing $\sum_{i=1}^n (i / (n + 1) - F_{\textrm{GE}}(X_{(i)}; \lambda, \nu))^2$. Similarly, in the Weighted Least Square (WLS) estimation, we obtain $\hat{\lambda}_{\textrm{WLS}}$ and $\hat{\nu}_{\textrm{WLS}}$, the estimators of $\lambda$ and $\nu$, by minimizing $\sum_{i=1}^n w_i (i / (n + 1) - F_{\textrm{GE}}(X_{(i)}; \lambda, \nu))^2$, where $w_i = (n + 1)^2 (n + 2) / (i (n - i + 1))$. Here, higher weights are assigned to both the tails of the sampling distribution. Further details are in \cite{gupta2001generalized}.

In the $L$-moments (LM) estimation, instead of obtaining the parameter estimates by equating the raw/central population moments and the sample moments as in the MM estimation, we equate the population moments and the sample moments of some linear combinations of order statistics. Because of using order statistics, this estimation approach is considered to be more robust compared to the traditional MM estimation. For the GE distribution, \cite{gupta2001generalized} use the estimating equations
\begin{eqnarray}
\nonumber (\psi(\nu + 1) - \psi(1)) / \lambda &=& n^{-1} \sum_{i=1}^n X_{(i)} = \Bar{X}, \\
\nonumber (\psi(2\nu + 1) - \psi(\nu + 1)) / \lambda &=& 2 \sum_{i=1}^n (i-1) X_{(i)} / (n (n - 1)) - \Bar{X},
\end{eqnarray}
to obtain the LM estimators $\hat{\lambda}_{\textrm{LM}}$ and $\hat{\nu}_{\textrm{LM}}$ of $\lambda$ and $\nu$, respectively. We compare the performance of these approaches with MDPDE through simulation studies.


\subsection{The minimum density power divergence estimation}

In MDPDE, we obtain the parameter estimates by minimizing DPD $d_\alpha(g,f)$ in (\ref{dpd}), over the parameter space. It involves a tuning parameter $\alpha\geq 0$, and for $\alpha = 0$, DPD is the same as the KL divergence. For the GE distribution, suppose the parameter vector is denoted by $\bm{\theta} = (\lambda, \nu)' \in \Theta = \mathbb{R}^{+} \times \mathbb{R}^{+}$. For parameter estimation, DPD is calculated as $d_\alpha(g,f_{\textrm{GE}}(\cdot; \bm{\theta}))$, where $g$ denotes the true density function and $f_{\textrm{GE}}$ is as given in (\ref{density_ge}). As previously mentioned, MLE is a special case of MDPDE when $\alpha = 0$, i.e., when DPD coincides with the KL divergence. Because MLE is highly sensitive to outliers, KL divergence is not suitable for robust parameter estimation, and thus choosing an appropriate divergence like DPD is important. Here we define the minimum DPD functional (bivariate) $\bm{T}_\alpha(\cdot)$ by \citep{basu1998robust}
\begin{equation}\label{EQ:dpd2}
d_\alpha(g, f_{\textrm{GE}}(\cdot; \bm{T}_\alpha(G))) = \min_{\bm{\theta} \in \Theta} d_\alpha(g,f_{\textrm{GE}}(\cdot; \bm{\theta})),
\end{equation}
where $G$ denotes the true distribution and $g$ is the density function of $G$. Thus, $\bm{T}_\alpha(G)$ is the parameter vector that provides the best fit under $G$. In practice, however, the true density $g$ is never known, and hence to obtain the minimizer of DPD, we need to use an estimate of $g$. Unlike other robust divergence measures, the particular DPD family proposed by \cite{basu1998robust} and given in (\ref{dpd}) has a major advantage because of not requiring any nonparametric smoothing for estimating $g$. This is because we can rewrite the DPD $d_\alpha(g,f_{\textrm{GE}}(\cdot; \bm{\theta}))$ as 
\begin{eqnarray}\label{EQ:dpd3}
 && d_\alpha(g,f_{\textrm{GE}}(\cdot; \bm{\theta})) \\
\nonumber     &=&\displaystyle \left\{\begin{array}{ll}
    \displaystyle \int f^{1+\alpha}_{\textrm{GE}}(x; \bm{\theta}) dx - \left(1 + \alpha^{-1}\right) E \left[ f^{\alpha}_{\textrm{GE}}(X; \bm{\theta}) \right] + 
\alpha^{-1} E\left[ g^{\alpha}(X)\right] & {\rm if} ~\alpha > 0,\\
	\displaystyle E\left[\log g(X)\right] - E\left[\log f_{\textrm{GE}}(X; \bm{\theta})\right] & {\rm if} ~\alpha = 0,  
\end{array}\right.
\end{eqnarray}
where $E\left[ \cdot \right]$ denotes the expectation with respect to $g$. Here, the two terms $E\left[g^\alpha(X) \right]$ and $E\left[\log g(X)\right]$ are independent of $\bm{\theta}$, and hence for the optimization in (\ref{EQ:dpd2}), they can be ignored. The other two expectations $E \left[ f^{\alpha}_{\textrm{GE}}(X; \bm{\theta}) \right]$ and $E\left[\log f_{\textrm{GE}}(X; \bm{\theta})\right]$ in (\ref{EQ:dpd3}) can be estimated directly through empirical means and does not require any nonparametric smoothing for estimating $g$. 
Therefore, the final estimator of $\bm{\theta}$ in MDPDE for tuning parameter $\alpha$ is defined as
\begin{equation}\label{EQ:dpd4}
\widehat{\bm{\theta}}_\alpha = \arg \min_{\bm{\theta} \in \Theta} H_{\alpha, n}(\bm{\theta}),
\end{equation}
where $H_{\alpha, n}(\bm{\theta}) = n^{-1} \sum_{i=1}^n V_\alpha(\bm{\theta}; X_i)$
with
\begin{equation}\label{EQ:dpd5}
   V_\alpha(\bm{\theta}; x)  = \displaystyle \left\{\begin{array}{ll}
    \displaystyle \int f^{1+\alpha}_{\textrm{GE}}(x; \bm{\theta}) dx - \left(1 + \alpha^{-1}\right) f^{\alpha}_{\textrm{GE}}(x; \bm{\theta}),  & {\rm ~~if} ~\alpha > 0,\\
	\displaystyle - \log f_{\textrm{GE}}(x; \bm{\theta}), & {\rm ~~if} ~\alpha = 0.  
\end{array}\right.
\end{equation}
For $\alpha = 0$, $\widehat{\bm{\theta}}_0 = \arg \min_{\bm{\theta} \in \Theta} n^{-1} \sum_{i=1}^n [- \log f_{\textrm{GE}}(X_i; \bm{\theta})]$
is clearly same as the ML estimator $\hat{\bm{\theta}}_{\textrm{ML}} = (\hat{\lambda}_{\textrm{ML}}, \hat{\nu}_{\textrm{ML}})'$, by definition. For implementing the optimization routine in (\ref{EQ:dpd4}), first we need to derive the explicit expression of $V_\alpha(\bm{\theta}; x)$ for the GE distribution. For $\alpha = 0$, it can be easily obtained from (\ref{density_ge}) and for $\alpha > 0$, it is given by
\begin{eqnarray}\label{EQ:dpd5_simple}
\nonumber V_\alpha(\bm{\theta}; x) &=& \lambda^\alpha \nu^\alpha \left[ \nu \textsc{B}(1 + \alpha, 1 + (1 + \alpha)(\nu - 1)) \right. \\
 &&~~~~~~~~ \left. - (1 + \alpha^{-1}) \exp[-\alpha \lambda x] (1 - \exp[-\lambda x])^{\alpha(\nu - 1)} \right],
\end{eqnarray}
where $\textsc{B}(a,b)$ denotes the beta function with inputs $a$ and $b$.

\section{Robustness and asymptotic properties}
\label{theory}

Robustness properties of MDPDE for the GE distribution directly follow from recognizing the estimator as an $M$-estimator, that is, the estimator satisfies an equation $\sum_{i=1}^n\bm{\psi}(X_i, \bm{\theta}) = \bm{0}$, where $\bm{\psi}(\cdot)$ is a vector-valued function of dimension same as that of the parameter space; for the GE distribution, it is thus two-dimensional. Here, unbiased estimating equations can be obtained for any $\alpha\geq 0$ through the differentiation of $H_{\alpha, n}(\bm{\theta})$ in (\ref{EQ:dpd4}) and the respective equations for the GE parameters $\lambda$ and $\nu$ are given by 
\begin{eqnarray}\label{EQ:dpd6}
\nonumber U_n(\lambda; \nu) \equiv \frac{1}{n} \sum_{i=1}^{n} u_{\lambda; \nu} (X_i)   f^{\alpha}_{\textrm{GE}}(X_i; \bm{\theta}) - \int u_{\lambda; \nu}(x) f^{1+\alpha}_{\textrm{GE}}(x; \bm{\theta}) dx = 0,\\
U_n(\nu; \lambda) \equiv \frac{1}{n} \sum_{i=1}^{n} u_{\nu; \lambda}(X_i) f^{\alpha}_{\textrm{GE}}(X_i; \bm{\theta}) - \int u_{\nu; \lambda}(x) f^{1+\alpha}_{\textrm{GE}}(x; \bm{\theta}) dx = 0,
\end{eqnarray}
where $u_{\lambda; \nu} (x) = \partial \log f_{\textrm{GE}}(x; \bm{\theta}) / \partial \lambda$ and $u_{\nu; \lambda}(x) = \partial \log f_{\textrm{GE}}(x; \bm{\theta}) / \partial \nu$ are the score functions.
At $\alpha=0$, (\ref{EQ:dpd6}) boils down to the usual estimating equations for MLE, while for any $\alpha>0$, the MDPDE gives weighted score equations with weights $f^{\alpha}_{\textrm{GE}}(X_i; \bm{\theta})$ for $X_i$. For any outlier $X_i$ with respect to the GE distribution, the evaluated density $f_{\textrm{GE}}(X_i; \bm{\theta})$ would be small and thus MDPDE produces robust estimates by down-weighting the effects of outliers in the estimating equations.

Further, we attempt to simplify (\ref{EQ:dpd6}) to obtain analytical expressions of the involved terms. The score vector $\bm{u}_{\bm{\theta}}(x) = (u_{\lambda; \nu} (x), u_{\nu; \lambda}(x))'$ is given by
\begin{equation} \label{score_vector}
    \bm{u}_{\bm{\theta}}(x) = 
\begin{pmatrix} 
\lambda^{-1} - x + (\nu - 1)x\exp[-\lambda x] / (1 - \exp[-\lambda x]) \\ 
\nu^{-1} + \log(1 - \exp[-\lambda x]) 
\end{pmatrix},
\end{equation}
and the terms $U_n(\lambda; \nu)$ and $U_n(\nu; \lambda)$ in (\ref{EQ:dpd6}) are given by 
\begin{eqnarray}
\nonumber U_n(\lambda; \nu) &=& \frac{1}{n} \sum_{i=1}^{n} \left\lbrace (\lambda^{-1} - X_i + (\nu - 1)X_i\exp[-\lambda X_i] / (1 - \exp[-\lambda X_i])) \right. \\
\nonumber && ~~~~~~~~~\left. \times~ \lambda^\alpha \nu^\alpha \exp[-\alpha \lambda X_i] (1 - \exp[-\lambda X_i])^{\alpha(\nu - 1)} \right \rbrace \\
\nonumber && - \lambda^{\alpha - 1} \nu^{\alpha + 1} \textsc{B}(1 + \alpha, (1 + \alpha) (\nu - 1) + 1) (1 +  \psi(1 + \alpha) - \psi(2 + \alpha)), \\
\nonumber U_n(\nu; \lambda) &=& \frac{1}{n} \sum_{i=1}^{n} \left\lbrace (\nu^{-1} + \log(1 - \exp[-\lambda X_i])) \right. \\
\nonumber && ~~~~~~~~~\left. \times~ \lambda^\alpha \nu^\alpha \exp[-\alpha \lambda X_i] (1 - \exp[-\lambda X_i])^{\alpha(\nu - 1)} \right \rbrace \\
 && - \lambda^{\alpha} \nu^{\alpha} \textsc{B}(1 + \alpha, (1 + \alpha) (\nu - 1) + 1) \\
\nonumber && ~~~~~~~~~ \times (1 + \nu (\psi((1 + \alpha) (\nu - 1) + 1) - \psi((1 + \alpha) \nu + 1)));
\end{eqnarray}
these simplifications require the condition $\nu > \alpha / (1 + \alpha)$ for any $\alpha \geq 0$.

\subsection{Asymptotic relative efficiency}

We further explore some theoretical properties of the estimator $\hat{\bm{\theta}}_{\alpha}$ in (\ref{EQ:dpd4}). The closed form expression of $\hat{\bm{\theta}}_{\alpha}$ is not available, even for the case of MLE ($\alpha = 0$). Similarly, the closed-form expressions for the mean and variance of $\hat{\bm{\theta}}_{\alpha}$ do not exist as well. For measuring the correctness of the estimator, exploring these properties is important. Despite the exact sampling distribution of $\hat{\bm{\theta}}_{\alpha}$ being difficult to find, the asymptotic results follow from \cite{basu1998robust}. Suppose, we assume that the model is correctly specified so that the true data generating distribution function is $G= F_{\textrm{GE}}(\cdot; \bm{\theta}^\ast)$ for $\bm{\theta}^\ast = (\lambda^\ast, \nu^\ast)'\in \Theta$. Then, under certain regularity conditions, \cite{basu1998robust} show that
$\widehat{\bm{\theta}}_\alpha = (\widehat{\lambda}_{\alpha}, \widehat{\nu}_{\alpha})'$ is a consistent estimator of $\bm{\theta}$ and the asymptotic distribution of $\tilde{\bm{\theta}}_{\alpha, n} = \sqrt{n} (\widehat{\bm{\theta}}_\alpha - \bm{\theta}^\ast)$ is bivariate normal with mean vector $\bm{\mu}_{\tilde{\bm{\theta}}} = (0,0)'$ and covariance matrix
$\bm{\Sigma}_{\tilde{\bm{\theta}}} = \bm{J}_\alpha(\bm{\theta}^\ast)^{-1}\bm{K}_\alpha(\bm{\theta}^\ast) \bm{J}_\alpha(\bm{\theta}^\ast)^{-1}$, 
where
\begin{eqnarray}\label{EQ:J_K_xi}
\nonumber && \bm{J}_\alpha(\bm{\theta}) = \int \bm{u}_{\bm{\theta}}(x) \bm{u}_{\bm{\theta}}(x)' f^{1+\alpha}_{\textrm{GE}}(x; \bm{\theta}) dx, \\
\nonumber && \bm{K}_\alpha(\bm{\theta}) = \bm{J}_{2\alpha}(\bm{\theta}) - \bm{\xi}_{\alpha}(\bm{\theta}) \bm{\xi}_{\alpha}(\bm{\theta})',~ \textrm{where}~ \\
&& \bm{\xi}_{\alpha}(\bm{\theta}) = 
\int \bm{u}_{\bm{\theta}}(x) f^{1+\alpha}_{\textrm{GE}}(x; \bm{\theta}) dx.
\end{eqnarray}

\begin{figure}[h]
    \centering
\includegraphics[width = \linewidth]{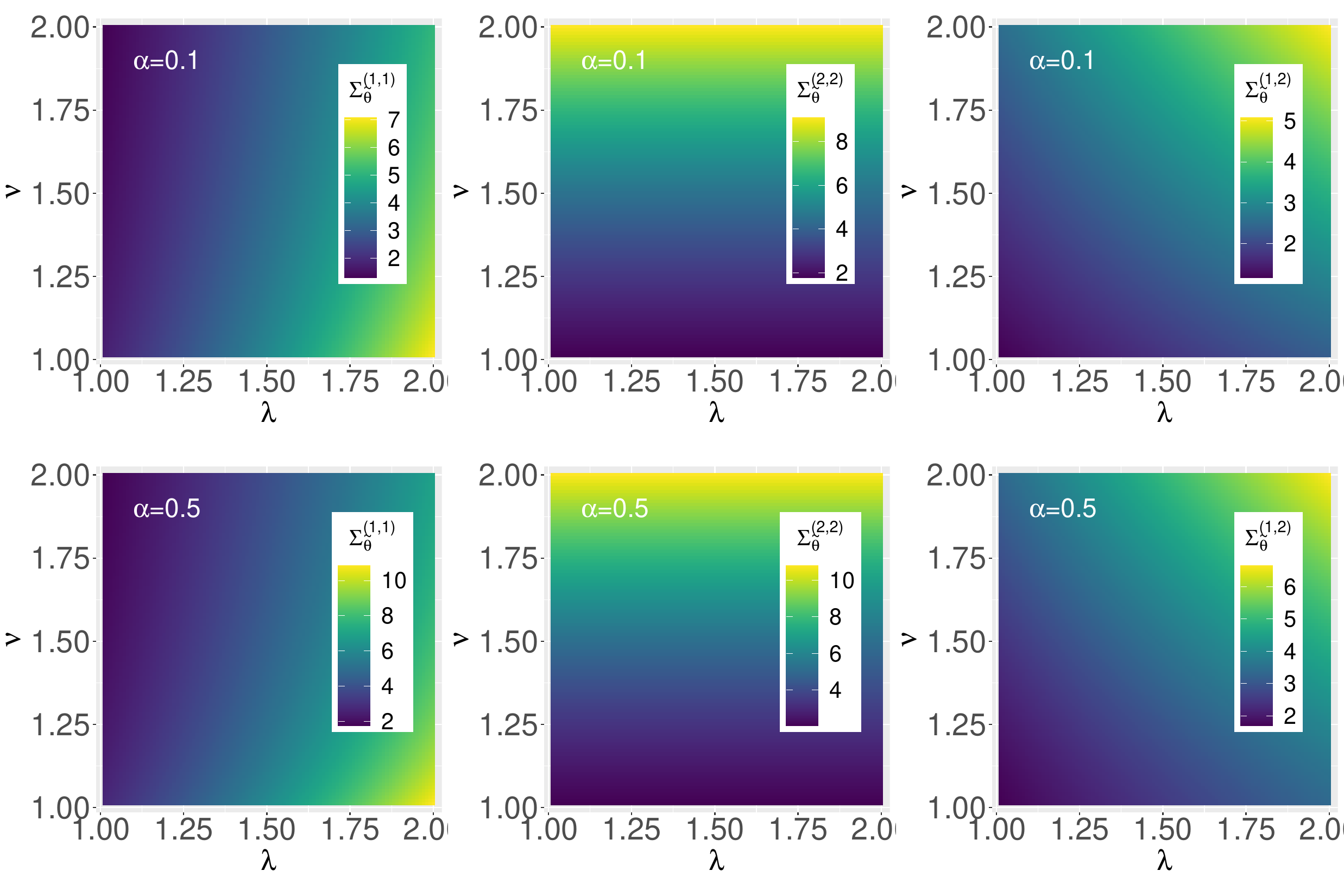}
    \caption{Variation of the elements of the covariance matrix $\bm{\Sigma}_{\tilde{\bm{\theta}}}$ for different values of $\lambda$, $\nu$, and $\alpha$. Here $\bm{\Sigma}^{(i,j)}_{\tilde{\bm{\theta}}}$ denotes the $(i,j)$-th element of $\bm{\Sigma}_{\tilde{\bm{\theta}}}$ for $i,j=1,2$.}
    \label{fig_ad_cov_comparison}
\end{figure}

Further, we attempt to simplify the expressions in (\ref{EQ:J_K_xi}). Let the $(i,j)$-th element of the symmetric matrix $\bm{J}_\alpha(\bm{\theta})$ be denoted by $\bm{J}^{(i,j)}_\alpha(\bm{\theta})$ for $i,j=1,2$. The vector $\bm{\xi}_{\alpha}(\bm{\theta})$ and the elements of $\bm{J}_\alpha(\bm{\theta})$ are given by
\begin{eqnarray}\label{EQ:J_K_xi_simple}
\nonumber &&\bm{\xi}_{\alpha}(\bm{\theta}) = \lambda^{\alpha - 1} \nu^\alpha\textsc{B}(1 + \alpha, (1 + \alpha) (\nu - 1) + 1) \\
\nonumber &&~~~~ \times \begin{pmatrix} 
\vspace{3mm} 
\nu (1 +  \psi(1 + \alpha) - \psi(2 + \alpha)) \\
\lambda (1 + \nu (\psi((1 + \alpha) (\nu - 1) + 1) - \psi((1 + \alpha) \nu + 1)))
\end{pmatrix}, \\ \nonumber \\
\nonumber && \bm{J}^{(1,1)}_\alpha(\bm{\theta}) = \lambda^{\alpha - 2} \nu^{\alpha+1}\textsc{B}(1 + \alpha, (1 + \alpha) (\nu - 1) + 1) \\ \nonumber &&~~ \times \left\lbrace \psi'(1 + \alpha) - \psi'(\nu(1 + \alpha) + 1) + (1 + \psi(1 + \alpha) - \psi(\nu(1 + \alpha) + 1))^2 \right. \\
\nonumber &&~~ \left. -2\left[\psi'(2 + \alpha) - \psi'(\nu(1 + \alpha) + 1) + \psi(2 + \alpha) - \psi(\nu(1 + \alpha) + 1) \right. \right.\\
\nonumber &&~~ \left. \left. + (\psi(2 + \alpha) - \psi(\nu(1 + \alpha) + 1))^2\right] + (\nu - 1)(\alpha + 2) / ((\nu - 1)(1 + \alpha) - 1) \right. \\
\nonumber &&~~\left. \times [\psi'(3 + \alpha) - \psi'(\nu(1 + \alpha) + 1) + (\psi(3 + \alpha) - \psi(\nu(1 + \alpha) + 1))^2] \right\rbrace, \\\nonumber \\
\nonumber && \bm{J}^{(2,2)}_\alpha(\bm{\theta}) = \lambda^\alpha \nu^{\alpha-1}\textsc{B}(1 + \alpha, (1 + \alpha) (\nu - 1) + 1) \\ \nonumber &&~~ \times \left\lbrace \nu^2 (\psi'((1 + \alpha)(\nu - 1) + 1) - \psi'(\nu(1 + \alpha) + 1)) \right. \\
\nonumber &&~~ \left. + (1 + \nu  (\psi((1 + \alpha)(\nu - 1) + 1) - \psi(\nu(1 + \alpha) + 1)))^2 \right\rbrace, \\\nonumber \\
\nonumber && \bm{J}^{(1,2)}_\alpha(\bm{\theta}) = \bm{J}^{(2,1)}_\alpha(\bm{\theta}) = \lambda^{\alpha - 1} \nu^\alpha\textsc{B}(1 + \alpha, (1 + \alpha) (\nu - 1) + 1)  \\ \nonumber &&~~ \times \left\lbrace 1 + \psi(\alpha + 1) - \psi(\alpha + 2) + \nu [(\psi((1 + \alpha) (\nu - 1) + 1) - \psi(\nu(1 + \alpha) + 1)) \right. \\
\nonumber &&~~ \times (1 + \psi(1 + \alpha) - \psi(\nu(1 + \alpha) + 1)) - (\psi(\alpha + 2) - \psi(\nu(1 + \alpha) + 1)) \\
 &&~~\left. \times (\psi((1 + \alpha) (\nu - 1)) - \psi(\nu(1 + \alpha) + 1)) ] \right\rbrace;
\end{eqnarray}
these simplifications require the condition $\nu > 1$ for any $\alpha \geq 0$, except for $\nu = (2 + \alpha)/(1 + \alpha)$. For $\alpha = 0$, the corresponding expressions are discussed in \cite{gupta2001exponentiated}. The final expressions for the elements of $\bm{\Sigma}_{\tilde{\bm{\theta}}}$ are functions of the model parameters $\lambda$ and $\nu$ and the MDPDE tuning parameter $\alpha$. We plot $\bm{\Sigma}^{(i,j)}_{\tilde{\bm{\theta}}}$, the $(i,j)$-th element of $\bm{\Sigma}_{\tilde{\bm{\theta}}}$ for $i,j=1,2$, for different values of $\lambda$, $\nu$, and $\alpha$ in Figure \ref{fig_ad_cov_comparison}. The term $\bm{\Sigma}^{(1,1)}_{\tilde{\bm{\theta}}}$, the asymptotic variance of $\sqrt{n}\widehat{\lambda}_{\alpha}$, is large when $\lambda$ is high, $\nu$ is low, and $\alpha$ is high, while $\bm{\Sigma}^{(2,2)}_{\tilde{\bm{\theta}}}$, the asymptotic variance of $\sqrt{n}\widehat{\nu}_{\alpha}$, is independent of $\lambda$ and large when $\nu$ is high and $\alpha$ is high. The off-diagonal element $\bm{\Sigma}^{(1,2)}_{\tilde{\bm{\theta}}}$ (same as $\bm{\Sigma}^{(2,1)}_{\tilde{\bm{\theta}}}$), the asymptotic covariance between $\sqrt{n}\widehat{\lambda}_{\alpha}$ and $\sqrt{n}\widehat{\nu}_{\alpha}$ is large when $\lambda$, $\nu$, and $\alpha$ are high. Theoretically, we obtain the following proportionality relations.

\begin{theorem} \label{thm1}
For the asymptotic covariance matrix $\bm{\Sigma}_{\tilde{\bm{\theta}}}$ of the rescaled estimator $\widehat{\bm{\theta}}_{\alpha}$, the relations between the elements of $\bm{\Sigma}_{\tilde{\bm{\theta}}}$ and the scale parameter $\lambda$ are as follows: (i) $\bm{\Sigma}^{(1,1)}_{\tilde{\bm{\theta}}} \propto \lambda^2$, (ii) $\bm{\Sigma}^{(1,2)}_{\tilde{\bm{\theta}}} \propto \lambda$, and (iii) $\bm{\Sigma}^{(2,2)}_{\tilde{\bm{\theta}}}$ is independent of $\lambda$.

\begin{proof}
Plugging the expressions in (\ref{EQ:J_K_xi_simple}) to (\ref{EQ:J_K_xi}), the matrices $\bm{J}_{\alpha}(\bm{\theta})$ and $\bm{K}_{\alpha}(\bm{\theta})$ are of the forms
\begin{equation}
\nonumber \bm{J}_{\alpha}(\bm{\theta}) = \begin{pmatrix} 
\vspace{3mm} 
\lambda^{\alpha - 2} J_{1,1}(\bm{\phi}) & \lambda^{\alpha - 1} J_{1,2}(\bm{\phi}) \\
\lambda^{\alpha - 1} J_{2,1}(\bm{\phi}) & \lambda^{\alpha} J_{2,2}(\bm{\phi})
\end{pmatrix}, \bm{K}_{\alpha}(\bm{\theta}) = \begin{pmatrix} 
\vspace{3mm} 
\lambda^{2\alpha - 2} K_{1,1}(\bm{\phi}) & \lambda^{2\alpha - 1} K_{1,2}(\bm{\phi}) \\
\lambda^{2\alpha - 1} K_{2,1}(\bm{\phi}) & \lambda^{2\alpha} K_{2,2}(\bm{\phi}),
\end{pmatrix}
\end{equation}
where $\bm{\phi} = (\nu, \alpha)'$ and $J_{i,j}(\cdot)$'s and $K_{i,j}(\cdot)$'s are functions of $\bm{\phi}$, independent of $\lambda$. Further, straightforward calculations show that the elements of $\bm{\Sigma}_{\tilde{\bm{\theta}}} = \bm{J}_\alpha(\bm{\theta})^{-1}\bm{K}_\alpha(\bm{\theta}) \bm{J}_\alpha(\bm{\theta})^{-1}$ can be written in the forms $\bm{\Sigma}^{(1,1)}_{\tilde{\bm{\theta}}} = \lambda^2 L_{1,1}(\bm{\phi})$, $\bm{\Sigma}^{(1,2)}_{\tilde{\bm{\theta}}} = \lambda L_{1,2}(\bm{\phi})$ and $\bm{\Sigma}^{(2,2)}_{\tilde{\bm{\theta}}} = L_{2,2}(\bm{\phi})$, where $L_{i,j}(\cdot)$'s are functions of $\bm{\phi}$, can be expressed in terms of $J_{i,j}(\cdot)$'s and $K_{i,j}(\cdot)$'s and are independent of $\lambda$.

\end{proof}

\end{theorem}

One can verify that the asymptotic variance of the estimator in MDPDE is minimum when $\alpha = 0$, i.e., in the case of MLE. Among two unbiased estimators, we generally prefer an estimator that has a smaller variance, and hence, we need to study the asymptotic relative efficiency (ARE), the ratio of the asymptotic variance of the estimator in MLE over that in MDPDE, assuming no outlier is present in the dataset. If ARE is close to one, the standard errors of the estimators for both methods are comparable, and hence, robustness can be achieved with only a small compromise in uncertainty. By definition, the ARE of the estimator in MDPDE is one if $\alpha=0$. From (\ref{EQ:J_K_xi_simple}), it is evident that the matrices $\bm{J}_\alpha(\bm{\theta})$ and $\bm{K}_\alpha(\bm{\theta})$ and hence $\bm{\Sigma}_{\tilde{\bm{\theta}}}$ can be simplified only up to in terms of polygamma functions. After studying the variation of ARE of the estimators in MDPDE with respect to $\lambda$, $\nu$, and $\alpha$, we obtain the following theorem.

\begin{theorem} \label{thm2}
The ARE of the estimators $\widehat{\lambda}_\alpha$ and $\widehat{\nu}_\alpha$ are independent of $\lambda$.

\begin{proof}
From Theorem \ref{thm1}, we have the asymptotic variance terms $\bm{\Sigma}^{(1,1)}_{\tilde{\bm{\theta}}} = \lambda^2 L_{1,1}(\bm{\phi})$ and $\bm{\Sigma}^{(2,2)}_{\tilde{\bm{\theta}}} = L_{2,2}(\bm{\phi})$, where $L_{i,j}(\cdot)$'s are functions of $\bm{\phi} = (\nu, \alpha)'$ and are independent of $\lambda$. Here,
$\textrm{ARE}(\widehat{\lambda}_\alpha) = \lambda^2 L_{1,1}((\nu, 0)') / (\lambda^2 L_{1,1}((\nu, \alpha)')) = L_{1,1}((\nu, 0)') / L_{1,1}((\nu, \alpha)')$ and thus independent of $\lambda$. Also, $\textrm{ARE}(\widehat{\nu}_\alpha) = L_{2,2}((\nu, 0)') / L_{2,2}((\nu, \alpha)'))$ is naturally independent of $\lambda$.
\end{proof}
\end{theorem}

In Table \ref{fig_are_comparison}, we present $\textrm{ARE}(\widehat{\lambda}_\alpha)$ and $\textrm{ARE}(\widehat{\nu}_\alpha)$ for different values of $\nu$ and $\alpha$. Here we observe that both $\textrm{ARE}(\widehat{\lambda}_\alpha)$ and $\textrm{ARE}(\widehat{\nu}_\alpha)$ decrease as $\alpha$ increases; however, the loss is not high for low values of $\alpha$. Thus, under the no outlier scenario, the asymptotic variances of the estimators in MDPDE are comparable with those in MLE at least for smaller $\alpha$ values, and thus, in case MDPDE provides high robustness against outlier (discussed in the next subsection), it is justified to use MDPDE over MLE. As $\alpha$ increases, $\textrm{ARE}(\widehat{\lambda}_\alpha)$ drops faster than $\textrm{ARE}(\widehat{\nu}_\alpha)$ for small values of $\nu$. Further analyses (not shown) show that for any fixed $\alpha > 0$, $\textrm{ARE}(\widehat{\lambda}_\alpha)$ is monotonically increasing with $\nu$, while $\textrm{ARE}(\widehat{\nu}_\alpha)$ shows a non-monotonic behavior and is low for small and large $\nu$ and $\arg\max_{\nu}\{\textrm{ARE}(\widehat{\nu}_\alpha)\}$ decreases monotonically as $\alpha$ increases.

\begin{table}[h]
\begin{center}
    \caption{\normalsize{Variation of $\textrm{ARE}(\widehat{\lambda}_\alpha)$ and $\textrm{ARE}(\widehat{\nu}_\alpha)$ for different combinations of $\nu$ and $\alpha$.}} \label{fig_are_comparison}
\begin{tabular}{|c |cccc |cccc |}
\hline
                             & \multicolumn{4}{c}{$\textrm{ARE}(\widehat{\lambda}_\alpha)$} & \multicolumn{4}{c}{$\textrm{ARE}(\widehat{\nu}_\alpha)$} \vline \\
                             \hline
\backslashbox{$\alpha$}{$\nu$}  & 1.25   & 2.5   & 5   & 10      & 1.25    & 2.5          & 5     & 10         \\
\hline
0.1            & 0.97          & 0.97          & 0.97          & 0.98          & 0.99         & 0.98         & 0.98         & 0.98         \\
0.2              & 0.90          & 0.92          & 0.92          & 0.92          & 0.96         & 0.95         & 0.94         & 0.93         \\
0.3              & 0.82          & 0.85          & 0.86          & 0.86          & 0.93         & 0.90         & 0.88         & 0.87         \\
0.4                & 0.74          & 0.78          & 0.79          & 0.80          & 0.89         & 0.86         & 0.82         & 0.81         \\
0.5              & 0.68          & 0.73          & 0.74          & 0.74          & 0.86         & 0.81         & 0.77         & 0.75         \\
0.6             & 0.62          & 0.68          & 0.69          & 0.69          & 0.84         & 0.77         & 0.72         & 0.70         \\
0.7             & 0.58          & 0.64          & 0.65          & 0.65          & 0.81         & 0.73         & 0.68         & 0.66         \\
0.8               & 0.55          & 0.60          & 0.62          & 0.62          & 0.79         & 0.70         & 0.65         & 0.62         \\
0.9              & 0.52          & 0.58          & 0.59          & 0.59          & 0.77         & 0.67         & 0.61         & 0.59         \\
1.0               & 0.49          & 0.55          & 0.57          & 0.57          & 0.75         & 0.64         & 0.59         & 0.56     \\
\hline
\end{tabular}
\end{center}
\end{table} 

\subsection{Influence function analysis}
\label{sec:influence}

Here we study the robustness of the estimators in MDPDE through the widely used influence function proposed by \citep{Hampeletc:1986}. Let $G_{\pi,x} =(1-\pi)G + \pi \delta_x$ be the contaminated distribution function where $G$ is the true data distribution function without any outlier, $\pi$ is the outlier proportion, and $\delta_x$ is the degenerate distribution at an outlier $x$.
Then, for the DPD functional $\bm{T}_{\alpha}(\cdot)$ in (\ref{EQ:dpd2}), $\left(\bm{T}_\alpha(G_{\pi,x}) - \bm{T}_\alpha(G)\right)$ is the (asymptotic) bias of $\widehat{\bm{\theta}}_\alpha$ due to the presence of an outlier. The influence function (IF) represents the standardized asymptotic bias of a robust estimator due to an infinitesimally small proportion of contamination and is given by
\begin{eqnarray}\label{EQ:if_def}
\textrm{IF}(x; \bm{T}_\alpha, G) = \lim_{\pi\rightarrow 0}\frac{\bm{T}_\alpha(G_{\pi,x}) - \bm{T}_\alpha(G)}{\pi}.
\end{eqnarray}
In our case, the GE distribution is parameterized by two parameters and thus $\textrm{IF}: \mathbb{R} \rightarrow \mathbb{R}^2$. If the function IF is not bounded, the bias of the underlying estimator can be very high even for an infinitesimally small proportion of contamination and thus the corresponding estimator would be non-robust. On the other hand, if the function IF remains bounded over $x$, the estimator would be within a bounded neighborhood of the true estimator $\bm{T}_\alpha(G)$ even under contamination at an extremely outlying observation $x$. Thus, to study the robustness of the estimator $\widehat{\bm{\theta}}_\alpha$, we study the boundedness of the function IF. 

When the model is correctly specified, i.e., $G= F_{\textrm{GE}}(\cdot; \bm{\theta}^\ast)$ for $\bm{\theta}^\ast = (\lambda^\ast, \nu^\ast)'\in \Theta$, following \cite{basu1998robust},
the function IF of the MDPDE functional $\bm{T}_\alpha(\cdot)$ is
\begin{eqnarray}\label{EQ:if_mdpde}
\nonumber \textrm{IF}(x; \bm{T}_\alpha, F_{\textrm{GE}}(\cdot; \bm{\theta}^\ast)) &=& \bm{J}_\alpha(\bm{\theta}^\ast)^{-1} \bigg[ \bm{u}_{\bm{\theta}^\ast}(x) f^{\alpha}_{\textrm{GE}}(x; \bm{\theta}^\ast) \\
&&~~~~~~~~~~~~~ \left. - \int \bm{u}_{\bm{\theta}^\ast}(y) f^{1+\alpha}_{\textrm{GE}}(y; \bm{\theta}^\ast) dy\right],
\end{eqnarray}
where $\bm{u}_{\bm{\theta}^\ast}(\cdot)$ is as defined in (\ref{score_vector}), and $\bm{J}_\alpha(\cdot)$ is as defined in (\ref{EQ:J_K_xi}) and (\ref{EQ:J_K_xi_simple}).

\begin{theorem}
Suppose the true value of the GE parameters $\lambda^\ast$ and $\nu^\ast$ are finite and $\nu^\ast > 1$ so that the derivations in (\ref{EQ:J_K_xi_simple}) are valid. Then, each component of the bivariate function $\textrm{IF}(x; \bm{T}_\alpha, F_{\textrm{GE}}(\cdot; \bm{\theta}^\ast))$ in (\ref{EQ:if_mdpde}) is bounded if and only if $\alpha > 0$.

\begin{proof}
Here $\bm{J}_\alpha(\bm{\theta}^\ast)^{-1}$ and $\int \bm{u}_{\bm{\theta}^\ast}(y) f^{1+\alpha}_{\textrm{GE}}(y; \bm{\theta}^\ast)dy$ within $\textrm{IF}(x; \bm{T}_\alpha, F_{\textrm{GE}}(\cdot; \bm{\theta}^\ast))$ are independent of $x$. Thus, only the vector $\bm{u}_{\bm{\theta}^\ast}(x) f^{\alpha}_{\textrm{GE}}(x; \bm{\theta}^\ast)$ determines the boundedness. Because of each component of $\textrm{IF}(x; \bm{T}_\alpha, F_{\textrm{GE}}(\cdot; \bm{\theta}^\ast))$ is a linear combination (weights are given by the rows of $\bm{J}_\alpha(\bm{\theta}^\ast)^{-1}$) of the elements in $\bm{u}_{\bm{\theta}^\ast}(x) f^{\alpha}_{\textrm{GE}}(x; \bm{\theta}^\ast)$, both the elements in $\bm{u}_{\bm{\theta}^\ast}(x) f^{\alpha}_{\textrm{GE}}(x; \bm{\theta}^\ast)$ are required to be bounded for $\textrm{IF}(x; \bm{T}_\alpha, F_{\textrm{GE}}(\cdot; \bm{\theta}^\ast))$ being bounded. 

First, we prove the `only if' part. Suppose, we assume that $\textrm{IF}(x; \bm{T}_\alpha, F_{\textrm{GE}}(\cdot; \bm{\theta}^\ast))$ is bounded but still $\alpha = 0$. Thus, each component of $\bm{u}_{\bm{\theta}^\ast}(x)$ is also bounded. However, the first component of $\bm{u}_{\bm{\theta}^\ast}(x)$ is given by $u_{\lambda^\ast, \nu^\ast}(x) = \lambda^{\ast -1} - x + (\nu^* - 1) x \exp[-\lambda^* x]/ (1 - \exp[-\lambda^* x])$ and $u_{\lambda^\ast, \nu^\ast}(x)$ is unbounded with $\lim_{x \rightarrow \infty} u_{\lambda^\ast, \nu^\ast}(x) = -\infty$. Similarly, the second component of $\bm{u}_{\bm{\theta}^\ast}(x)$ is given by $u_{\nu^\ast, \lambda^\ast}(x) = \nu^{\ast -1} + \log(1 - \exp[-\lambda^* x])$ and $u_{\nu^\ast,\lambda^\ast}(x)$ is unbounded with $\lim_{x \rightarrow 0} u_{\nu^\ast,\lambda^\ast}(x) = -\infty$. Hence, the contradiction.

Further, we prove the `if' part. The first component of $\bm{u}_{\bm{\theta}^\ast}(x) f^{\alpha}_{\textrm{GE}}(x; \bm{\theta}^\ast)$ is clearly a continuous function defined over $\mathbb{R}^+$. Thus, for any $0 < a_m < x < b_m < \infty$, the component $u_{\lambda^\ast, \nu^\ast}(x) f^{\alpha}_{\textrm{GE}}(x; \bm{\theta}^\ast)$ is bounded over $[a_m, b_m]$ follows from the extreme value theorem in calculus. Further, considering sequences $a_m \rightarrow 0$ and $b_m \rightarrow \infty$, the boundedness is also valid for any $m$ and it is easy to show that $\lim_{x \rightarrow 0} u_{\lambda^\ast, \nu^\ast}(x) f^{\alpha}_{\textrm{GE}}(x; \bm{\theta}^\ast) = \lim_{x \rightarrow \infty} u_{\lambda^\ast, \nu^\ast}(x) f^{\alpha}_{\textrm{GE}}(x; \bm{\theta}^\ast) = 0$. Thus, $u_{\lambda^\ast, \nu^\ast}(x) f^{\alpha}_{\textrm{GE}}(x; \bm{\theta}^\ast)$ is bounded over $\mathbb{R}^+$. Using a similar approach, we can also show that the second component of $\bm{u}_{\bm{\theta}^\ast}(x) f^{\alpha}_{\textrm{GE}}(x; \bm{\theta}^\ast)$ is also bounded and it is again easy to show that $\lim_{x \rightarrow 0} u_{\nu^\ast,\lambda^\ast}(x) f^{\alpha}_{\textrm{GE}}(x; \bm{\theta}^\ast) = \lim_{x \rightarrow \infty} u_{\nu^\ast, \lambda^\ast}(x) f^{\alpha}_{\textrm{GE}}(x; \bm{\theta}^\ast) = 0$.
\end{proof}
\end{theorem}

\begin{figure}[]
	\centering
	\includegraphics[width = 0.8\linewidth]{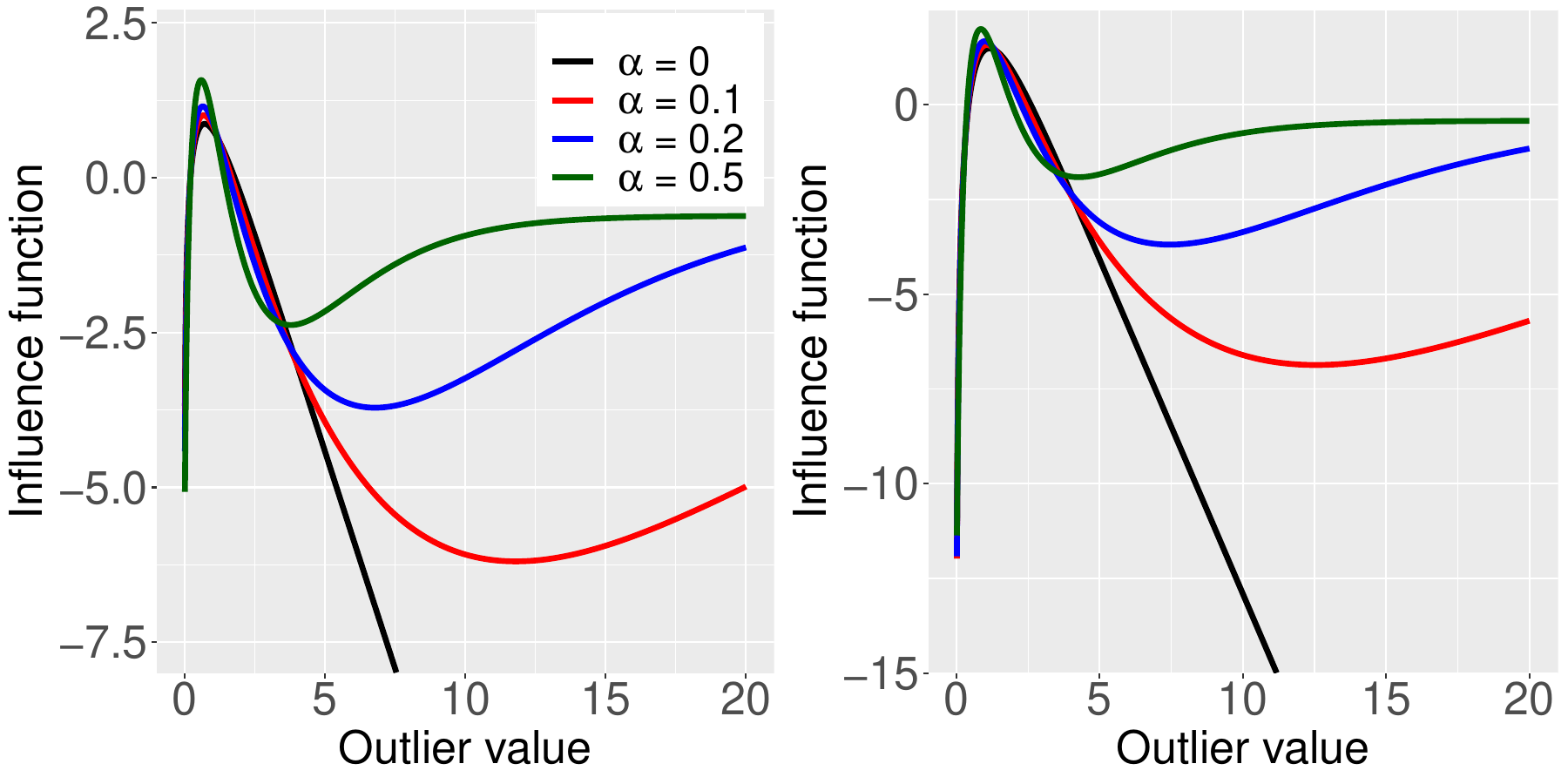}
\caption{Components of the influence function $\textrm{IF}(x; \bm{T}_\alpha, F_{\textrm{GE}}(\cdot; \bm{\theta}^\ast))$ for $\lambda^\ast$ (left panel) and $\nu^\ast$ (right panel) at $\textrm{GE}(1, 1.5)$ distribution.}
	\label{fig_if}
\end{figure}

To visualize the boundedness, we illustrate $\textrm{IF}(x; \bm{T}_\alpha, F_{\textrm{GE}}(\cdot; \bm{\theta}^\ast))$ in Figure \ref{fig_if}, for different choices of $\alpha$, over the outlier value $x$, for the GE distribution with specific (true values) model parameters $\lambda^\ast$ and $\nu^\ast$. Here we choose the true parameters to be $\lambda^\ast = 1$ and $\nu^\ast = 1.5$. The boundedness of the components of $\textrm{IF}(x; \bm{T}_\alpha, F_{\textrm{GE}}(\cdot; \bm{\theta}^\ast))$ at $\alpha>0$ and their unboundedness at $\alpha=0$ are clear from Figure \ref{fig_if}. When $\alpha = 0.5$, the components of the influence function stabilize faster compared to the cases when $\alpha = 0.1$ or $\alpha = 0.2$. Thus, based on the influence function analysis, the robustness of the estimators in MDPDE is confirmed, while we observe a non-robust behavior in the case of MLE (at $\alpha=0$).

\subsection{Optimal tuning parameter selection}
\label{optimal_alpha}
The tuning parameter $\alpha$ in MDPDE provides a trade-off between the robustness and efficiency of the underlying estimators and thus $\alpha$ needs to be appropriately chosen depending on the proportion of outliers present in the data. A high value of $\alpha$ should be chosen when the proportion of outliers is high and vice versa. However, the outlier proportion is not known in practice and thus a data-driven algorithm is required for selecting an optimal $\alpha$.
 Here we follow the approach proposed by \cite{fujisawa2006robust} and opted by \cite{seo2017extreme} and \cite{hazra2019robust},
where the optimal $\alpha$ is chosen by minimizing the empirical Cramer-von Mises (CVM) distance as
\begin{equation}\label{EQ:CVM}
\alpha_{\textrm{opt}} = \arg \min_{\alpha} \frac{1}{n} \sum_{i=1}^{n} \left\lbrace \frac{i}{n + 1} - F_{\textrm{GE}}(X_{(i)}; \widehat{\bm{\theta}}_\alpha^{(-i)}) \right\rbrace^2,
\end{equation}
where $X_{(1)}, \ldots, X_{(n)}$ are the order statistics and 
$\widehat{\bm{\theta}}_\alpha^{(-i)}$ is the estimator in MDPDE for tuning parameter $\alpha$ obtained after removing $X_i$ out from $\mathcal{X}$. Apart from (\ref{EQ:CVM}), several authors \citep{hong2001automatic, warwick2005choosing, basak2021optimal} have proposed other techniques for the tuning parameter selection as well and some of them are also applicable for any general $M$-estimator, not necessarily only for MDPDE.

\section{Simulation studies}
\label{simulation}

To compare the performances of MDPDE for different choices of the tuning parameter $\alpha$ including $\alpha_{\textrm{opt}}$ in \eqref{EQ:CVM}, along with the existing estimation strategies discussed in Section \ref{ge_parest}, we consider four different contamination or outlier proportions 0\%, 1\%, 5\%, and 10\%, respectively. We generate 1000 random samples of size $n=100$ from the $\textrm{GE}(1, 1.5)$ distribution (true data distribution) and replace different proportions (0\%, 1\%, 5\%, and 10\%) of the observations by two different choices of outliers. First (C1), we assume that the outliers follow a degenerate distribution supported at 7.31, the 0.999-th quantile or 1000-unit return level of the $\textrm{GE}(1, 1.5)$ distribution. While analyzing monthly or annual rainfall data for 100 years, observing one or a few data points more extreme than the 1000-year return level can be relatively common and such observations are crucial for extreme value analysis. However, when our focus is on the bulk of the distribution (i.e., we are interested in estimating the mean or median rainfall more accurately, for example), such observations can be treated as outliers. Second (C2), we assume that the outliers are degenerate at 14.22, the $(1 - 10^{-6})$-th quantile or $10^6$-unit return level of the $\textrm{GE}(1, 1.5)$ distribution. While analyzing 100 observations that are assumed to be identically distributed, one or a few data entries more extreme than the $10^6$-unit return level is unlikely and could have resulted from a wrong data entry or ill-functioning of the measuring instrument. Thus, the two simulation settings we consider here cover both possible scenarios of contamination. For each of the 1000 replications, we estimate the GE parameters using MDPDE and other methods in Section \ref{ge_parest}, and then calculate the average bias and mean square error (MSE) of the estimates. The results for C1 and C2 are presented in Table \ref{table_sim1} and \ref{table_sim2}, respectively.

\begin{table}[h]
\caption{\normalsize Average bias and mean square error (MSE) in the estimation of the GE parameters $\lambda$ and $\nu$ using different estimation methods under different outlier percentages. Here, outlier value is the $(1 - 10^{-3})$-th quantile of the true distribution $\textrm{GE}(1, 1.5)$. Highlighted entries in each column correspond to the cases with Bias and MSE closest to zero.}
\begin{tabular}{cccccccccc}
\hline
 &  & \multicolumn{8}{c}{Outlier percentage} \\
 &  & \multicolumn{2}{c}{0\%} & \multicolumn{2}{c}{1\%} & \multicolumn{2}{c}{5\%} & \multicolumn{2}{c}{10\%} \\
Method & $\bm{\theta}$ & Bias & MSE & Bias & MSE & Bias & MSE & Bias & MSE \\
\hline
\multirow{2}{*}{ML} & $\lambda$ & 0.021 & \textbf{0.016} & -0.054 & \textbf{0.015} & -0.278 & 0.081 & -0.445 & 0.199 \\
 & $\nu$ & 0.045 & \textbf{0.058} & -0.038 & \textbf{0.046} & -0.270 & 0.095 & -0.427 & 0.195 \\ \cline{2-10}
\multirow{2}{*}{MM} & $\lambda$ & 0.028 & 0.026 & -0.143 & 0.030 & -0.424 & 0.181 & -0.544 & 0.296 \\
 & $\nu$ & 0.080 & 0.138 & -0.248 & 0.109 & -0.629 & 0.408 & -0.699 & 0.498 \\ \cline{2-10}
\multirow{2}{*}{PT} & $\lambda$ & -0.052 & 0.029 & -0.226 & 0.062 & -0.493 & 0.244 & -0.569 & 0.324 \\
 & $\nu$ & -0.084 & 0.142 & -0.424 & 0.232 & -0.778 & 0.619 & -0.745 & 0.565 \\ \cline{2-10}
\multirow{2}{*}{LS} & $\lambda$ & \textbf{0.003} & 0.023 & -0.027 & 0.022 & -0.143 & 0.039 & -0.282 & 0.095 \\
 & $\nu$ & \textbf{0.013} & 0.080 & -0.018 & 0.075 & -0.135 & 0.078 & -0.276 & 0.121 \\ \cline{2-10}
\multirow{2}{*}{WLS} & $\lambda$ & 0.010 & 0.019 & -0.028 & 0.019 & -0.153 & 0.040 & -0.293 & 0.099 \\
 & $\nu$ & 0.024 & 0.066 & -0.016 & 0.060 & -0.139 & 0.068 & -0.271 & 0.111 \\ \cline{2-10}
\multirow{2}{*}{LM} & $\lambda$ & 0.011 & 0.019 & -0.096 & 0.021 & -0.363 & 0.135 & -0.523 & 0.274 \\
 & $\nu$ & 0.025 & 0.079 & -0.139 & 0.070 & -0.488 & 0.255 & -0.646 & 0.426 \\ \cline{2-10}
\multirow{2}{*}{\begin{tabular}[c]{@{}c@{}}MDPDE\\ ($\alpha = 0.1$)\end{tabular}} & $\lambda$ & 0.021 & 0.017 & -0.032 & 0.016 & -0.227 & 0.059 & -0.408 & 0.169 \\
 & $\nu$ & 0.045 & 0.058 & -0.012 & 0.049 & -0.211 & 0.072 & -0.383 & 0.161 \\ \cline{2-10}
\multirow{2}{*}{\begin{tabular}[c]{@{}c@{}}MDPDE\\ ($\alpha = 0.2$)\end{tabular}} & $\lambda$ & 0.023 & 0.018 & -0.014 & 0.017 & -0.169 & 0.041 & -0.353 & 0.131 \\
 & $\nu$ & 0.047 & 0.060 & \textbf{0.009} & 0.053 & -0.147 & 0.059 & -0.321 & 0.123 \\ \cline{2-10}
\multirow{2}{*}{\begin{tabular}[c]{@{}c@{}}MDPDE\\ ($\alpha = 0.5$)\end{tabular}} & $\lambda$ & 0.032 & 0.025 & 0.018 & 0.024 & -0.048 & \textbf{0.028} & -0.150 & 0.049 \\
 & $\nu$ & 0.058 & 0.071 & 0.045 & 0.068 & \textbf{-0.012} & 0.062 & -0.103 & \textbf{0.066} \\  \cline{2-10}
\multirow{2}{*}{\begin{tabular}[c]{@{}c@{}}MDPDE\\ ($\alpha = 1$)\end{tabular}} & $\lambda$ & 0.049 & 0.036 & 0.037 & 0.034 & \textbf{-0.016} & 0.031 & \textbf{-0.085} & \textbf{0.036} \\
 & $\nu$ & 0.082 & 0.096 & 0.073 & 0.091 & 0.032 & 0.079 & \textbf{-0.020} & 0.069
  \\ \cline{2-10}
\multirow{2}{*}{\begin{tabular}[c]{@{}c@{}}{MDPDE}\\ ($\alpha = \alpha_{\textrm{opt}}$)\end{tabular}} & $\lambda$ & 0.015 & 0.021 & \textbf{-0.013} & 0.020 & -0.127 & 0.032 & -0.257 & 0.078 \\
 & $\nu$ & 0.039 & 0.064 & 0.014 & 0.059 & -0.096 & \textbf{0.055} & -0.218 & 0.077 \\
 \hline
\end{tabular}
\label{table_sim1}
\end{table}

Table \ref{table_sim1} shows that under the true data scenario (0\% outlier), the average biases are positive for all the estimation methods except for PT estimation and both the parameters, while they are closest to zero for the LS estimation. The average biases for MDPDE with $\alpha = 0.1$ and ML estimation are similar and they increase as $\alpha$ increases in MDPDE. For $\alpha = \alpha_{\textrm{opt}}$, MDPDE returns a smaller average bias than ML estimation, while the MSE for MDPDE is higher than that for ML estimation. The two methods WLS and LM perform relatively better compared to ML estimation and MDPDE in terms of average bias. Despite higher biases, the ML method performs better than others in terms of MSE, and MDPDE with $\alpha = 0.1$ results in smaller MSE compared to the methods MM, PT, LS, WLS, and LM. When one out of $n=100$ observations is replaced with the outlier value for each replication, all methods except MDPDE with $\alpha = 0.5, 1$, return negative average biases, and for MDPDE with $\alpha = \alpha_{\textrm{opt}}$ and $\alpha = 0.2$, the average biases are closest to zero for the GE rate and shape parameters, respectively. In terms of MSE, the ML method still outperforms all the alternatives. While the MSE for MDPDE with $\alpha = \alpha_{\textrm{opt}}$ is larger than that for ML estimation, it is still smaller than that for other methods MM, PT, LS, WLS, and LM. When the outlier percentage is 5\%, MDPDE with different choices of $\alpha = 0.5, 1, \alpha_{\textrm{opt}}$ performs better than other methods. All the estimation methods lead to negative average biases for both $\lambda$ and $\nu$ except for MDPDE with $\alpha = 1$. Despite LM being claimed to be a robust estimation approach, it performs poorer than all other methods except MM and PT. For the outlier percentage 10\%, all the estimation methods and the choices of $\alpha$ considered here for MDPDE lead to negative biases for both $\lambda$ and $\nu$. MDPDE with high $\alpha = 1$ outperforms other methods as well as other choices of $\alpha$ in terms of the average bias. In terms of MSE, MDPDE with $\alpha = 1$ outperforms others for $\lambda$ and MDPDE with $\alpha = 0.5$ outperforms others for $\nu$. We also note that the MSE values for $\alpha = 0.5$ and $\alpha = 1$ are comparable. For $\alpha = \alpha_{\textrm{opt}}$, despite the MSE values for MDPDE being higher than those based on MDPDE with $\alpha=0.5, 1$, they are smaller than those based on all the competing methods. Thus, in a practical setting where the presence of one or more outliers in the data is suspected and the choice of $\alpha$ is arbitrary, choosing $\alpha = \alpha_{\textrm{opt}}$ is reasonable.

\begin{table}[t]
\caption{\normalsize Average bias and mean square error (MSE) in the estimation of the GE parameters $\lambda$ and $\nu$ using different estimation methods under different outlier percentages. Here, outlier value is the $(1 - 10^{-6})$-th quantile of the true distribution $\textrm{GE}(1, 1.5)$. Highlighted entries in each column correspond to the cases with Bias and MSE closest to zero.}
\begin{tabular}{cccccccccc}
\hline
 &  & \multicolumn{8}{c}{Outlier percentage} \\
 &  & \multicolumn{2}{c}{0\%} & \multicolumn{2}{c}{1\%} & \multicolumn{2}{c}{5\%} & \multicolumn{2}{c}{10\%} \\
Method & $\bm{\theta}$ & Bias & MSE & Bias & MSE & Bias & MSE & Bias & MSE \\
\hline
\multirow{2}{*}{ML} & $\lambda$ & 0.021 & \textbf{0.016} & -0.138 & 0.028 & -0.501 & 0.252 & -0.681 & 0.464 \\
 & $\nu$ & 0.045 & \textbf{0.058} & -0.146 & 0.055 & -0.554 & 0.316 & -0.741 & 0.552 \\ \cline{2-10}
\multirow{2}{*}{MM} & $\lambda$ & 0.028 & 0.026 & -0.454 & 0.207 & -0.745 & 0.555 & -0.808 & 0.653 \\
 & $\nu$ & 0.080 & 0.138 & -0.818 & 0.677 & -1.121 & 1.259 & -1.120 & 1.256 \\ \cline{2-10}
\multirow{2}{*}{PT} & $\lambda$ & -0.052 & 0.029 & -0.499 & 0.253 & -0.810 & 0.656 & -0.822 & 0.676 \\
 & $\nu$ & -0.084 & 0.142 & -0.926 & 0.876 & -1.263 & 1.596 & -1.150 & 1.324 \\ \cline{2-10}
\multirow{2}{*}{LS} & $\lambda$ & \textbf{0.003} & 0.023 & -0.027 & 0.022 & -0.142 & 0.038 & -0.273 & 0.089 \\
 & $\nu$ & \textbf{0.013} & 0.080 & -0.018 & 0.075 & -0.133 & 0.077 & -0.261 & 0.112 \\ \cline{2-10}
\multirow{2}{*}{WLS} & $\lambda$ & 0.010 & 0.019 & -0.027 & 0.019 & -0.147 & 0.037 & -0.269 & 0.085 \\
 & $\nu$ & 0.024 & 0.066 & -0.015 & 0.060 & -0.130 & 0.064 & -0.234 & 0.093 \\  \cline{2-10}
\multirow{2}{*}{LM} & $\lambda$ & 0.011 & 0.019 & -0.222 & 0.057 & -0.616 & 0.380 & -0.763 & 0.583 \\
 & $\nu$ & 0.025 & 0.079 & -0.346 & 0.149 & -0.855 & 0.736 & -1.003 & 1.009 \\ \cline{2-10}
\multirow{2}{*}{\begin{tabular}[c]{@{}c@{}}MDPDE\\ ($\alpha = 0.1$)\end{tabular}} & $\lambda$ & 0.021 & 0.017 & -0.040 & \textbf{0.017} & -0.316 & 0.109 & -0.594 & 0.353 \\
 & $\nu$ & 0.045 & 0.058 & -0.026 & \textbf{0.051} & -0.332 & 0.135 & -0.622 & 0.393 \\ \cline{2-10}
\multirow{2}{*}{\begin{tabular}[c]{@{}c@{}}MDPDE\\ ($\alpha = 0.2$)\end{tabular}} & $\lambda$ & 0.023 & 0.018 & \textbf{0.001} & 0.019 & -0.110 & 0.034 & -0.335 & 0.136 \\
 & $\nu$ & 0.047 & 0.060 & 0.024 & 0.057 & -0.094 & 0.063 & -0.329 & 0.152 \\ \cline{2-10}
\multirow{2}{*}{\begin{tabular}[c]{@{}c@{}}MDPDE\\ ($\alpha = 0.5$)\end{tabular}} & $\lambda$ & 0.032 & 0.025 & 0.026 & 0.024 & \textbf{-0.004} & \textbf{0.024} & \textbf{-0.045} & \textbf{0.026} \\
 & $\nu$ & 0.058 & 0.071 & 0.053 & 0.069 & \textbf{0.032} & 0.066 & \textbf{0.002} & \textbf{0.062} \\ \cline{2-10}
\multirow{2}{*}{\begin{tabular}[c]{@{}c@{}}MDPDE\\ ($\alpha = 1$)\end{tabular}} & $\lambda$ & 0.049 & 0.036 & 0.037 & 0.034 & -0.012 & 0.030 & -0.076 & 0.033 \\
 & $\nu$ & 0.082 & 0.096 & 0.073 & 0.091 & 0.036 & 0.079 & -0.010 & 0.068
 \\ \cline{2-10}
\multirow{2}{*}{\begin{tabular}[c]{@{}c@{}} MDPDE\\ ($\alpha = \alpha_{\textrm{opt}}$)\end{tabular}} & $\lambda$ & 0.015 & 0.021 & -0.013 & 0.020 & -0.097 & 0.030 & -0.187 & 0.065 \\
 & $\nu$ & 0.039 & 0.064 & \textbf{0.013} & 0.061 & -0.063 & \textbf{0.063} & -0.141 & 0.089 \\
 \hline
\end{tabular}
\label{table_sim2}
\end{table}

The first two columns of Table \ref{table_sim2} are naturally identical to those of Table \ref{table_sim1}. Once one of the $n=100$ observations is replaced with the outlier value for each replication, all methods except MDPDE with $\alpha = 0.2, 0.5, 1$ return negative average biases for both $\lambda$ and $\nu$. For MDPDE with $\alpha = \alpha_{\textrm{opt}}$, the average bias is negative for $\lambda$ and positive for $\nu$; however, both the values are close to zero. For MDPDE with $\alpha = 0.2$, the average bias is closest to zero for $\lambda$, and for $\nu$, MDPDE with $\alpha = \alpha_{\textrm{opt}}$ returns the least absolute average bias. Unlike the $1\%$ contamination scenario in C1, where ML estimation outperforms all the alternatives in terms of MSE, MDPDE with $\alpha = 0.1$ outperforms other methods in C2. When the outlier percentage is 5\%, MDPDE with different choices of $\alpha = 0.5, \alpha_{\textrm{opt}}$ performs better than other methods in terms of average bias and MSE. Similar to C1, the robust LM estimation approach again performs poorer than all other methods except MM and PT. For the outlier percentage 10\%, all the estimation methods lead to negative biases for both $\lambda$ and $\nu$, except for MDPDE with $\alpha = 0.5$ in the case of estimating $\nu$. In terms of both the average bias and MSE, MDPDE with $\alpha = 0.5$ outperforms all the alternative methods and other choices of $\alpha$. Similar to C1, in a practical setting, where the choice of $\alpha$ is arbitrary, choosing $\alpha = \alpha_{\textrm{opt}}$ is reasonable as it returns lower MSE and absolute bias compared to the competing methods (ML, MM, PT, LS, WLS, LM) in the presence of outliers.

The results from Table \ref{table_sim1} and \ref{table_sim2} demonstrate the effectiveness of MDPDE over other methods mentioned in Section \ref{ge_parest}, even in the presence of a small percentage of outliers in the dataset. When the contamination percentage is tiny (1\%) and the outlier value is smaller (C1), ML estimation still performs better than others in terms of MSE. However, once the outlier is larger (C2), MDPDE with some $\alpha > 0$ outperforms other alternatives. While an influence function analysis illustrates the limiting effect of an infinitesimally small proportion of outliers on estimation as explained in Section \ref{sec:influence}, the influence of a high contamination percentage can be assessed from Table \ref{table_sim1} and \ref{table_sim2}. When $\alpha > 0$, we observe a non-monotonic behavior of the influence functions in Figure \ref{fig_if}. Besides, for a certain range of values of the outlier, the influence function shows different increasing and decreasing patterns across different values of $\alpha$. Comparing the elements across Tables \ref{table_sim1} and \ref{table_sim2}, for $\alpha=0.1$, the absolute average biases in estimating $\lambda$ and $\nu$ increase as the outlier increases. At the same time, an opposite pattern is observable for $\alpha=0.2$. Comparing the absolute average biases for $\alpha = \alpha_{\textrm{opt}}$, we observe that they are uniformly smaller when the outlier is larger, although such a pattern is not observable in MSE. Overall, across different positive proportions of outliers, MDPDE with $\alpha = \alpha_{\textrm{opt}}$ generally outperforms the alternative methods.

\section{Rainfall data analysis}
\label{data_application}

\begin{figure}[]
	\centering
	\includegraphics[width = \linewidth]{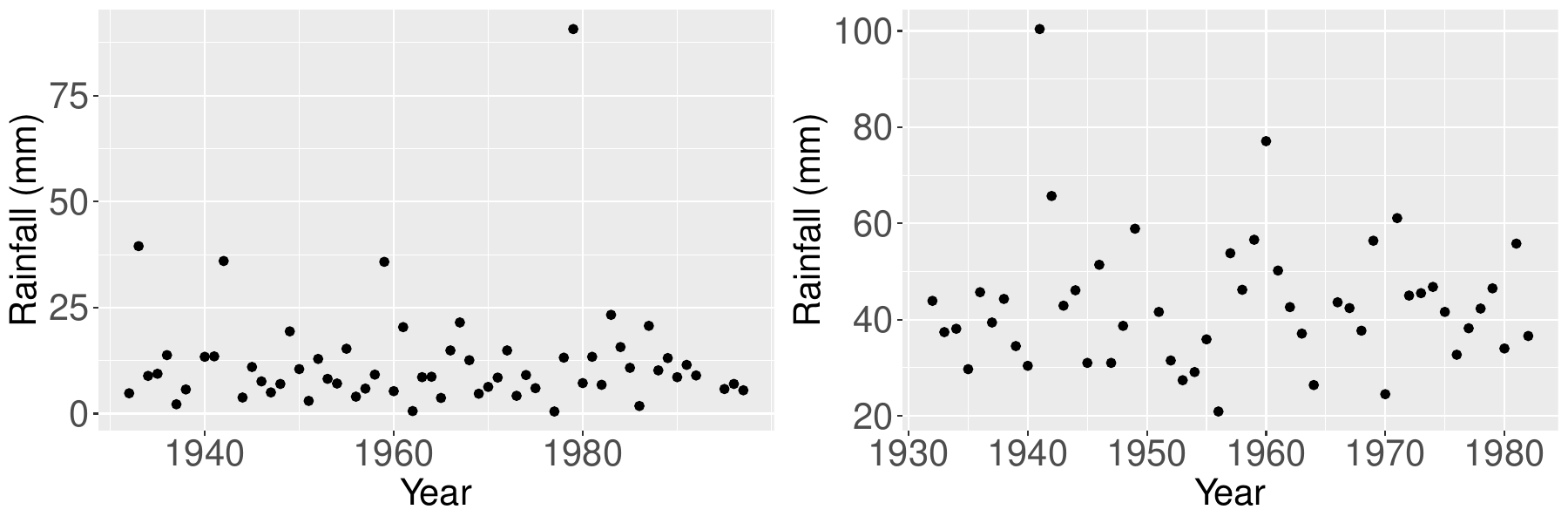}
\caption{Left: Time series of July rainfall at Brazoria, Texas, United States. Right: Time series of annual rainfall at Castro, Texas, United States.}
	\label{fig_tsplots}
\end{figure}

Here we analyze two rainfall datasets using MDPDE that illustrate the effectiveness of the robust estimation strategy discussed in this paper. Both datasets are provided by the Institute for Mathematics Applied to Geosciences (\url{https://www.image.ucar.edu/Data/US.monthly.met/}). First, we fit the GE distribution using MDPDE to the July rainfall dataset at Brazoria, Texas, United States. The observation period is between 1932--1997 and we have 61 observations available (other data are missing) across years only for July. Treating a month to be a short-period (thus high volatility), the chances of large outliers are higher. Second, annual rainfall data at Castro, Texas, United States, are analyzed using the same method. The observation period is between 1932--1982 and we have 50 observations available (other data are missing). Treating a year to be a long period (thus low volatility), the chances of observing extreme outliers are relatively lower compared to July rainfall. Both time series are presented in Figure \ref{fig_tsplots}. We call these two datasets `July dataset' and `Annual dataset', respectively.

We perform some exploratory analyses first to confirm the model assumptions: (i) there is no trend in the data so that we can safely assume that the observations are identically distributed, (ii) there is no strong temporal dependence in the data, (iii) GE distribution fits the data well after removing the outliers. For identifying trends, we use simple linear regression with the corresponding year as a covariate. For the two datasets, the $p$-values are 0.974 and 0.897, respectively. Thus, the observations in both datasets can be safely assumed to be identically distributed. Then we study the autocorrelation and partial autocorrelation functions (ACF and PACF, respectively). The lag-1 ACF (or PACF) for the two datasets are -0.0764 and 0.1127, respectively, and none of ACF and PACF appear to be significant for any of the two datasets for any of the first fifteen lags. Thus, the observations in both datasets can be safely assumed to be independent across time; despite the zero autocorrelation not indicating independence for a GE distributed time series, this exploratory analysis approach is widely used in spatial statistics literature for non-Gaussian models \citep{hazra2020multivariate, hazra2021estimating, hazra2021large}. For fitting the GE distribution to each dataset, first, we identify the outliers using a widely used adjusted boxplot method for skewed distributions proposed by \cite{hubert2008adjusted}. For estimating the model parameters, we use MDPDE for the full datasets (including outliers) and choose the optimal tuning parameter as described in Section \ref{optimal_alpha}. However, the outliers are treated as contamination in the pure GE distribution and thus naturally removed while performing a goodness-of-fit test for the GE distribution. We use the Kolmogorov-Smirnov (K-S) goodness-of-fit test to check the model fitting and obtain the $p$-values using a parametric bootstrap approach. The K-S test statistic and the corresponding $p$-value for the Monthly dataset are 0.0604 and 0.7890, respectively. For the Annual dataset, the K-S test statistic and the corresponding $p$-value are 0.1037 and 0.1786, respectively. Thus, we can safely assume that after removing the outliers, the observations are distributed as the GE distribution for both datasets.

\begin{figure}[h]
	\centering
	\includegraphics[width = \linewidth]{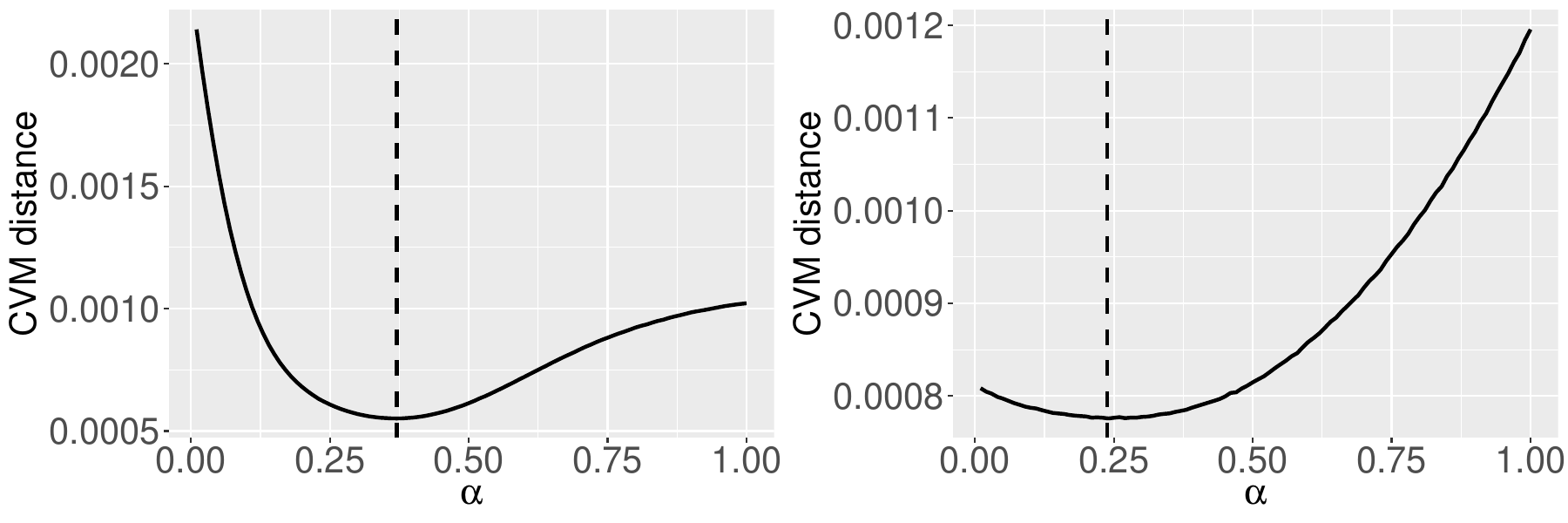}
\caption{MDPDE tuning parameter versus CVM distance in Section \ref{optimal_alpha} for July rainfall data at Brazoria, Texas, United States (left) and for annual rainfall at Castro, Texas, United States (right). The vertical dashed lines represent the optimal tuning parameter scenarios.}
	\label{fig_cvmdistplots}
\end{figure}

\begin{figure}[h]
	\centering
	\includegraphics[width = 0.49\linewidth]{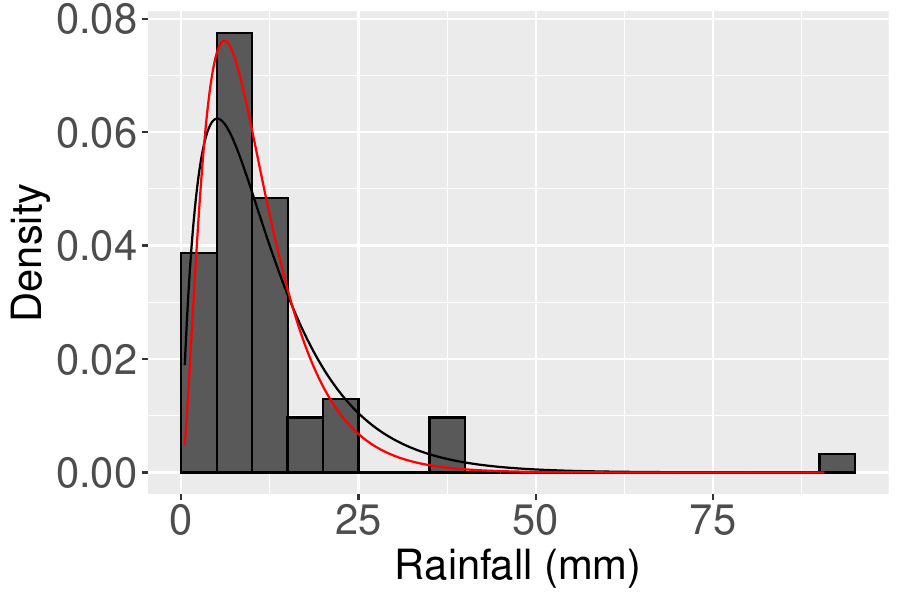}
	\includegraphics[width = 0.49\linewidth]{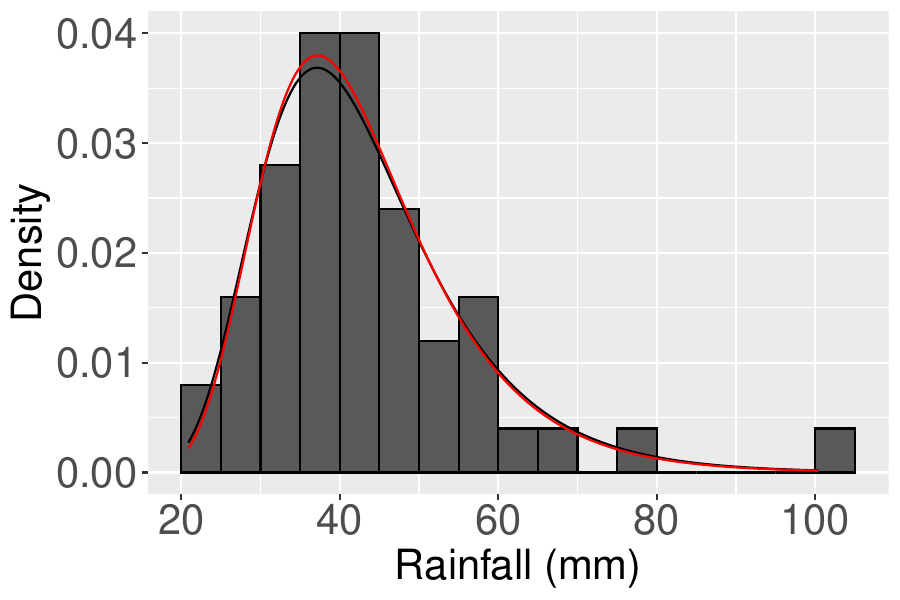}
\caption{Histograms of the July (left) and Annual (right) rainfall datasets overlapped with the fitted GE densities where the model parameters are estimated using ML estimation (black lines) and MDPDE with optimal tuning parameter values (red lines).}
	\label{fig_datavsfitsplots}
\end{figure}

Following Section \ref{optimal_alpha}, we further explore the CVM distances for both datasets for the tuning parameter $\alpha \in (0, 1)$ and $\alpha$ versus CVM distance curves are presented in Figure \ref{fig_cvmdistplots}. The optimal $\alpha$ choices for the Monthly and Annual datasets are 0.3705 and 0.2374, respectively. The most influential outlier (for the year 1979) present in the Monthly dataset is more extreme than the most influential outlier (for the year 1941) present in the Annual dataset and a higher optimal $\alpha$ for the Monthly dataset reflects the necessity for more robust parameter estimation compared to Annual dataset. The fitted GE densities for both datasets using MDPDE with optimal $\alpha$ and ML estimation are presented in Figure \ref{fig_datavsfitsplots}. For the Monthly dataset, ML estimation is highly influenced by an extreme outlier and as a result, it underestimates the density near the bulk of the data, while the robust MDPDE estimates the density near the bulk of the data more accurately. For the annual dataset, the outlier is not extremely high and thus do not influence the parameter estimation heavily. As a result, MDPDE with optimal $\alpha$ performs only slightly better than ML estimation; the mode of the MDPDE-based fitted density is closer to the mode of the empirical density (histogram) compared to that based on ML estimation.

Further, we estimate the GE parameters and the K-S distances between empirical and fitted GE distributions based on different estimation procedures described in Section \ref{ge_parest} for the July and Annual datasets, and the values are reported in Table \ref{table_parests_real}. For MDPDE, the tuning parameter is chosen as $\alpha_{\textrm{opt}}$ in (\ref{EQ:CVM}). For the July dataset, the estimated rate parameter $\hat{\lambda}$ is small for each method ranging between 0.0617 for PT estimation through 0.1724 for LS estimation. Except for MM and PT estimation methods, all other approaches return the estimated shape parameter $\hat{\nu}$ to be more than one. The histogram in the left panel of Figure \ref{fig_datavsfitsplots} clearly shows a non-monotonically decreasing behavior of the data density and thus an estimated $\nu$ larger than one is more justified. While MDPDE outperforms other alternatives (in terms of K-S distance), LS and WLS methods also perform well in terms of model fitting. For the Annual dataset, the estimates $\hat{\lambda}$'s are again small for each method ranging between 0.0679 for LM estimation through 0.1052 for LS estimation. However, the variation of the estimates across different methods is relatively smaller for the Annual dataset compared to the July dataset. The estimates $\hat{\nu}$'s for the Annual dataset are way larger than those for the July dataset. A larger $\nu$ indicates smaller skewness (bottom-right panel of Figure \ref{fig_ge_properties}) and smaller volatility and skewness for the Annual rainfall is expected due to the longer observational period compared to July rainfall. Here LS estimation outperforms other alternatives in terms of K-S distance; however, the performance of MDPDE is also comparable. Specifically, in comparison with the mostly used ML estimation, $D_{\textrm{KS}}$ for MDPDE is smaller for both datasets, with a large improvement being prominent for the July dataset. Overall, similar to the results based on simulation studies, we observe that MDPDE performs slightly better than the alternative models in the presence of extremely large outliers while the method performs comparable with other approaches when the proportion of outliers is small and the outlier values are only moderately large.

\begin{table}[]
\caption{\normalsize Estimated GE parameters ($\hat{\lambda}, \hat{\nu}$) and the K-S distance between empirical and fitted GE distributions ($D_{\textrm{KS}}$) based on different estimation procedures for the July and Annual datasets. For MDPDE, the tuning parameter is chosen as $\alpha_{\textrm{opt}}$ in (\ref{EQ:CVM}). The highlighted entries correspond to the cases with the smallest $D_{\textrm{KS}}$.}
\centering
\begin{tabular}{cccclccc}
\hline
 & \multicolumn{3}{c}{July dataset} &  & \multicolumn{3}{c}{Annual dataset} \\ \cline{2-4} \cline{6-8}
 & $\hat{\lambda}$ & $\hat{\nu}$ & $D_{\textrm{KS}}$ &  & $\hat{\lambda}$ & $\hat{\nu}$ & $D_{\textrm{KS}}$ \\
 \hline 
ML & 0.1208 & 1.8484 & 0.1145 &  & 0.0989 & 39.4420 & 0.1006 \\
MM & 0.0755 & 0.8613 & 0.2157 &  & 0.0926 & 29.9146 & 0.1066 \\
PT & 0.0617 & 0.5975 & 0.3064 &  & 0.0867 & 23.8370 & 0.1197 \\
LS & 0.1724 & 2.9162 & 0.0584 &  & 0.1052 & 50.9216 & \textbf{0.0911} \\
WLS & 0.1665 & 2.7709 & 0.0585 &  & 0.1017 & 43.7221 & 0.0938 \\
LM & 0.1083 & 1.5447 & 0.1224 &  & 0.0679 & 9.9999 & 0.1449 \\
MDPDE & 0.1683 & 2.7775 & \textbf{0.0578} &  & 0.1021 & 44.6385 & 0.0943 \\
\hline
\end{tabular}
\label{table_parests_real}
\end{table}

\section{Discussions and conclusions}
\label{conclusion}

The generalized exponential (GE) distribution has been widely used in several disciplines including rainfall modeling over the last two decades. However, as of the author's knowledge, it has been used for rainfall modeling only rarely, while some competing models like exponential, gamma, log-normal, and Weibull distributions are predominantly used. Rainfall datasets often exhibit a positive skewness which resembles the theoretical properties of the GE distribution. Thus, we propose modeling rainfall data using the GE distribution in the agro-meteorology literature. 

There are several existing parameter estimation techniques for the GE distribution in the literature; they are maximum likelihood (ML) estimation, method of moments estimation, percentile estimation, least square estimation, and $L$-moment estimation. However, such techniques do not ensure robustness against strong outliers often present in short-period rainfall datasets and also common in long-period (e.g., annual) rainfall datasets. ML estimation is the most common parameter estimation strategy in both statistical as well as meteorological literature due to some of its attractive theoretical properties and the availability of software implementing the procedure for the practitioners. However, considering its sensitivity to outliers, here we discuss a robust parameter estimation procedure, namely the minimum density power divergence estimation (MDPDE) proposed by \cite{basu1998robust}. In MDPDE, we obtain the estimates by minimizing a density-based divergence measure. In this paper, we derive the analytical expressions for the estimating equations and asymptotic distributions of the estimators. Then, using influence function analysis, we compare MDPDE with ML estimation analytically in terms of robustness and prove the boundedness of the influence functions for MDPDE. Besides, we also study the asymptotic relative efficiency of MDPDE analytically for different GE shape parameters and MDPDE tuning parameters and show that for a moderate tuning parameter value, the asymptotic relative efficiency is close to one. Further, we apply the proposed method to simulated datasets as well as to two rainfall datasets from Texas, United States. The results indicate a better performance of MDPDE compared to the other techniques in most of the scenarios, specifically when some high outliers are present in the dataset. 

While our main goal in this paper is establishing the theoretical details for MDPDE for the GE parameters and implementing the method for analyzing univariate rainfall data, usually the precipitation profiles are studied across different monitoring stations leading to the requirement of multivariate or spatial modeling. Multivariate or spatial modeling has been studied widely in the spatial statistics literature \citep{cressie1993statistics}, where superior estimates can be obtained by borrowing information from nearby sites. A multivariate GE distribution and a stationary GE stochastic process are developed by \cite{kundu2015absolute} and \cite{kundu2021stationary}, respectively. Analyzing multivariate or spatial rainfall data using them would be a future endeavor. \cite{basu1998robust} originally proposed MDPDE for univariate data; however, a multivariate version of it has been recently studied by \cite{chakraborty2020robust}. Implementing MDPDE in a spatial setting is a possible future research direction. 
One possible direction would be a two-stage analysis where the GE distributions can be fitted to rainfall datasets separately at each station, and then spatially smooth parameter surfaces can be obtained using the tools in spatial statistics; one such example is the max-and-smooth approach discussed in \citep{johannesson2021approximate, hazra2021large} and MDPDE can also be readily used in the first stage in such a scenario.


\section*{Supplementary information}
Codes (written in \texttt{R}) for obtaining MDPDEs, their standard errors, and optimal tuning parameter for the generalized exponential distribution are provided in the Supplementary Material (also available at \url{https://github.com/arnabstatswithR/robustgenexp.git}).

\section*{Acknowledgement}
We would like to thank the Editor-in-Chief and an anonymous reviewer for several thoughtful comments that improved the quality of the manuscript.

\section*{Disclosure statement}

No potential conflict of interest was reported by the author(s).

\section*{Data availability statement}

Datasets used in this paper are provided by the Institute for Mathematics Applied to Geosciences available at \url{https://www.image.ucar.edu/Data/US.monthly.met/}.

\bibliographystyle{agsm}
\bibliography{sn-bibliography}

\end{document}